\documentstyle[preprint,eqsecnum,aps,epsf]{revtex}
\newif\iftightenlines\tightenlinesfalse
\tightenlines\tightenlinestrue

\def\eslt{\not\!\!{E_T}}
\def\to{\rightarrow}

\def\tu{\tilde u}

\def\tb{\tilde b}

\def\tst{\tilde t}
\def\ttau{\tilde \tau}

\def\tg{\tilde g}

\def\tq{\tilde q}
\def\tw{\widetilde W}

\def\tz{\widetilde Z}
\begin{document}
\draft
\preprint{\vbox{\baselineskip=14pt%
   \rightline{FSU-HEP-000216}
   \rightline{UCCHEP/12-00}
   \rightline{IFIC/00-18}
   \rightline{FTUV/000504}
   \rightline{UH-511-963-00}
}}
\title{YUKAWA UNIFIED
SUPERSYMMETRIC SO(10) MODEL: \\
COSMOLOGY, RARE DECAYS AND COLLIDER SEARCHES}
\author{Howard Baer$^1$, Michal Brhlik$^2$, Marco A. D\'\i az$^{3}$, 
Javier Ferrandis$^{4}$, \\ Pedro Mercadante$^1$, Pamela Quintana$^1$
and Xerxes Tata$^5$}
\address{
$^1$Department of Physics,
Florida State University,
Tallahassee, FL 32306 USA
}
\address{
$^2$Randall Physics Laboratory
University of Michigan,
Ann Arbor, MI 48109-1120 USA
}
\address{
$^3$Facultad de F\'\i sica, Universidad Cat\'olica de Chile,
Av. Vicu\~na Mackenna 4860, Santiago, Chile
}
\address{
$^4$Departament de F\'\i sica Te\`orica,
Universitat de Val\`encia,
Spain
}
\address{
$^5$Department of Physics and Astronomy,
University of Hawaii,
Honolulu, HI 96822, USA
}
\date{\today}
\maketitle
\begin{abstract}

It has recently been pointed out that viable sparticle mass spectra 
can be generated in Yukawa unified $SO(10)$ supersymmetric grand
unified models consistent with radiative breaking of electroweak
symmetry. Model solutions are obtained only if $\tan\beta \sim 50$, 
$\mu <0$ and
positive $D$-term contributions to scalar masses from $SO(10)$ 
gauge symmetry breaking are used.
In this paper, we attempt to systematize the parameter space regions where
solutions are obtained. 
We go on to calculate the relic density of neutralinos as a function
of parameter space. No regions of the parameter space explored were 
actually cosmologically excluded, and very reasonable relic densities
were found in much of parameter space.
Direct neutralino detection rates could exceed 1 event/kg/day for a 
$^{73}$Ge detector, for low values of GUT scale gaugino mass $m_{1/2}$.
We also calculate the branching fraction for $b\to s \gamma$ decays, and find
that it is beyond the 95\% CL
experimental limits in much, but not all, of the parameter space regions
explored. However, recent claims have been made that NLO effects can reverse
the signs of certain amplitudes in the $b\to s\gamma$ calculation, leading to
agreement between theory and experiment in Yukawa unified SUSY models.
For the Fermilab Tevatron collider, significant regions of parameter
space can be explored via $b\bar{b}A$ and $b\bar{b}H$ searches.
There also exist some limited regions of parameter space
where a trilepton signal can be seen at TeV33. Finally, there exist
significant regions of parameter space where direct detection of
bottom squark pair production can be made, especially for large negative 
values of the GUT parameter $A_0$.

\end{abstract}

\medskip



\section{Introduction}

Supersymmetric $SO(10)$ grand unified theories\cite{so10} (GUTS) 
are highly motivated
choices for particle physics models which encompass and go 
beyond the Standard Model (SM). In particular:
\begin{itemize}
\item All three forces in the SM are unified by a simple Lie group. 
Since the model is supersymmetric, the hierarchy between weak and GUT
scales is naturally stabilized. 
\item The fifteen distinct fermions of each SM generation
plus a SM gauge singlet right-handed neutrino state are economically
included in the sixteen dimensional spinorial representation of $SO(10)$.
\item The group $SO(10)$ is anomaly-free, and provides an explanation for
cancellation of triangle anomalies that are otherwise rather ad-hoc in
both the MSSM and $SU(5)$ SUSY-GUT models.
\item The SM weak hypercharge assignments
can be derived from broken $SO(10)$ using only the 
lowest dimensional $SO(10)$ Higgs multiplets.
This has been interpreted as group theoretic evidence of
$SO(10)$ grand unification\cite{willenbrock}.
\item The singlet neutrino superfield(s) $\hat{N}^c$ may develop a Majorana
mass at very high energy scales $\sim 10^{11}-10^{16}$ GeV. 
Combining these with the usual Dirac neutrino masses that can now develop,
one is lead to (dominantly) left-handed neutrino masses at the eV scale (or 
below), while right handed neutrinos are pushed beyond observability,
via the see-saw mechanism\cite{seesaw}. The resulting neutrino masses
can easily be in accord with recent data from solar and atmospheric
neutrino detection experiments, though a detailed model is necessary for
an explanation of the observed mixing pattern.
\item In minimal $SO(10)$, there exists only a single Yukawa coupling
per generation, so that Yukawa couplings also unify at the $GUT$ scale.
This is especially constraining for the
third generation.  In addition, predictive $SO(10)$ models can be
created which encompass the fermion masses of the lighter generations as
well\cite{fmasses}.
\item In $SO(10)$ models, $R$-parity conservation is 
a natural consequence of the gauge symmetry of the model\cite{mohapatra}.
When $SO(10)$ is broken, if the breaking occurs via certain ``safe''
representations of the Higgs fields, $R$-parity is still conserved
even in the weak scale theory\cite{martin}. Along the same lines,
it has been shown that in SUSY models with gauged $B-L$ (this includes
$SO(10)$ models)
symmetry and a renormalizable see-saw mechanism, $R$-parity must be 
conserved exactly, even in the weak scale Lagrangian\cite{rasin}.
\item In $SO(10)$ models, baryogenesis in the early universe can be
explained as a consequence of decays of (out of thermal equilibrium)
intermediate scale right-handed neutrino states\cite{fy}.

\end{itemize}

Motivated by these considerations, in this paper we investigate a variety
of phenomenological implications of Yukawa-unified $SO(10)$ SUSY-GUT models.
Throughout our study, we assume the following:
\begin{enumerate}
\item A minimal $SO(10)$ SUSY GUT model with a single Yukawa coupling
per generation, valid at energy scales 
$Q>M_{GUT}\simeq 2\times 10^{16}$ GeV.
\item The gauge symmetry $SO(10)$ breaks via some (to be specified)
mechanism directly to the MSSM gauge group at $Q=M_{GUT}$. We assume
universality of soft SUSY breaking mass parameters at $Q=M_{GUT}$.
However, as a consequence of the reduction in rank of the 
gauge symmetry breaking,
additional $D$-term mass contributions are generated for scalar masses.
\item Electroweak symmetry is broken radiatively (REWSB).
\end{enumerate}
With these assumptions, the spectrum of 
supersymmetric particles and Higgs bosons can be calculated. 
We examine various phenomenological consequences of the sparticle mass
spectrum generated. We do not address the related question of mass
generation of first and second generation SM fermions; 
this issue has been investigated in a 
number of other papers\cite{fmasses}. We also do not explore the specific
mechanisms involved in gauge symmetry breaking; several studies along these
lines are contained in Ref. \cite{models}.

A crucial issue is to find models with Yukawa coupling unification
which reproduce the observed masses for the third generation fermions. One may 
assume Yukawa unification at $Q=M_{GUT}$, and evolve gauge and Yukawa
couplings and soft SUSY breaking (SSB) masses to the weak scale, 
and only accept solutions which generate third generation fermion masses
in agreement with measurements within tolerances. Alternatively, one may
begin with the central values of measured fermion masses at low energies, 
and accept only solutions
which give Yukawa unification to within set tolerances at the $GUT$ scale.
In this paper, we adopt the latter bottom-up approach.
A crucial aspect for this approach is to correctly calculate the weak 
scale third generation Yukawa couplings. 
Throughout our work, we 
fix pole masses $m_b=4.9$ GeV, $m_t=175$ GeV and $m_{\tau}=1.78$ GeV.

The mass spectrum of SUSY particles in minimal supersymmetric $SO(10)$ 
constrained by radiative electroweak symmetry breaking has been 
studied previously in a number of 
papers\cite{op1,barger,copw,shafi,cw,rsh,mn,hemp,strumia,op2,mop}. 
Unification of bottom, tau and top Yukawa couplings was found to occur at
very large values of the parameter $\tan\beta\sim 50-60$, and specific
spectra were generated for values of $m_t\sim 190$ GeV\cite{barger}.
Assuming universality of
soft SUSY breaking masses at $M_{GUT}$, it was found\cite{copw,cw} 
that Yukawa unification consistent with radiative electroweak
symmetry breaking could also occur for $m_t<170$ GeV as long as
$m_{1/2} \agt 300$ GeV. This generally leads to sparticle 
masses far beyond the reach of the CERN LEP2 or Fermilab Tevatron
$p\bar{p}$ colliders. 
For values of $m_t\simeq 175$ GeV, solutions including 
radiative electroweak breaking were very difficult to achieve.
In Ref. \cite{op2}, the SUSY particle mass spectrum
was investigated with {\it non-universal} SSB masses. Various solutions
were found, but ad hoc non-universality conflicts with the $SO(10)$
symmetry. In Ref. \cite{mop}, it was suggested that $SO(10)$ $D$-term
contributions to scalar masses had the correct form to allow for
successful radiative electroweak symmetry breaking and the computation
of weak scale SUSY particle masses, but explicit solutions were not 
presented.

In a previous paper\cite{bdft}, we have shown that Yukawa unified
solutions to the superparticle mass spectrum could be generated (even
for $m_t=175$~GeV) assuming universality within $SO(10)$, but only if
additional $D$-term contributions to scalar masses were included.
Solutions with Yukawa unification good to 5\% and REWSB were found only
for values of $\mu <0$ and positive values of the $D$-term
contributions.  A crucial ingredient in the calculation is the inclusion
of SUSY loop corrections to fermion masses\cite{rsh,copw,pierce}, from
which the third generation weak scale Yukawa couplings are
calculated. For $\mu >0$, the diminution in the $b$-quark Yukawa
coupling is so severe that not even $b-\tau$ unification can be acheived
for any values of $\tan\beta :2-60$. For values of $\tan\beta \agt 60$,
the Yukawa couplings diverge during renormalization group running from
the weak scale to the $GUT$ scale, so that these values of $\tan\beta$
form an upper limit to the allowed parameter space. For $\mu <0$,
assuming universality of SSB scalar masses, a much lower parameter space
limit of $\tan\beta \alt 45$ is found. For higher values of $\tan\beta$
we find $\mu^2<0$, and REWSB falters essentially because the Higgs mass
squared $m_{H_{d}}^2$ is driven to smaller weak scale values than
$m_{H_{u}}^2$.  However, if one allows positive $D$-term contributions
to scalar masses, then $m_{H_{u}}^2< m_{H_{d}}^2$ already at $M_{GUT}$,
and the region of allowed $\tan\beta$ expands to $\tan\beta\sim
50-55$. This is sufficient to allow for $t-b-\tau$ Yukawa unification
for many models with $\mu <0$. Of course, the $D$-term contributions
must be added to the various squark and slepton soft masses as well, so
that the $D$-terms leave a distinctive imprint on the superparticle mass
spectrum\cite{dterms}. The most striking possibility is that the light
bottom squark could be by far the lightest of all the squarks, and furthermore,
even within range of Fermilab Tevatron searches. In addition, left-
slepton masses could be lighter than right slepton masses, in
contrast to expectations from the SUSY models with universality at
$Q=M_{GUT}$.  This latter effect could certainly be extracted from
precision measurements at linear $e^+e^-$ colliders operating at
$\sqrt{s}\agt 1$ TeV, and would be evidence in favor of a Yukawa unified
$SO(10)$ SUSY GUT model.

In this paper, we address many of the observable aspects of
Yukawa-unified $SO(10)$ models in some detail.  In Sec. II, we summarize
the results from Ref. \cite{bdft}, and try to systematize the parameter
space expected in Yukawa unified $SO(10)$ model framework. In Sec. III,
we present calculations of the cosmological relic density of neutralinos
in this class of models. Over much of the parameter space, we find very
reasonable relic densities, unless $m_{\tz_1}\simeq {m_A\over 2}$, where
$s$-channel annihilation through a Higgs resonance can lower $\Omega
h^2$ to very tiny values.  Given a reasonable relic density, it is
well-known that rates for direct neutralino detection experiments are
highest at large $\tan\beta$. We show direct detection rates in $SO(10)$
parameter space, and show that much of the space should be explorable by
neutralino-nucleon scattering experiments.  In Sec. IV, we evaluate the
rate for $b\to s\gamma$ decays. Much of the parameter space is excluded
by the measured value of the branching fraction for the decay $b\to s
\gamma$, but some distinctive regions remain viable. These results may
be subject to modification if there is additional non-universality or
inter-generational squark mixing at $Q=M_{GUT}$, or if $CP$ violating
phases are significant. In addition, recent claims have been made that NLO
QCD corrections to $b\to s\gamma$ loop diagrams can reverse the signs of 
certain amplitudes, leading to {\it agreement} between experimental measurements
and theory predictions from Yukawa unified SUSY models. 
In Sec. V, we examine regions of parameter
space accessible to searches for supersymmetry at the Fermilab Tevatron
collider. We find three possibilities: {\it i}) signals from $Ab\bar{b}$
or $Hb\bar{b}$ associated production, {\it ii}) $\tw_1\tz_2\to 3\ell$
signals if $\tan\beta$ is not too big, and {\it iii}) signals from
direct production of $\tb_1\bar{\tb}_1$ pairs, where observable rates
occur for large negative values of $A_0$. Experiments at the Large
Hadron Collider (LHC) would, of course, decisively probe the model. We
end in Sec.~VI with a summary of our results, and some concluding
remarks.

\section{Minimal $SO(10)$ model parameter space}

In the minimal supersymmetric $SO(10)$ GUT model, the fifteen 
matter superfields of each generation of the MSSM plus a gauge singlet
neutrino superfield $\hat{N}^c$ are included in the 16-dimensional
spinorial representation $\hat{\psi}$ of $SO(10)$. 
In addition, the
two Higgs doublet superfields $\hat{H}_u$ and $\hat{H}_d$ are embedded
in a single 10-dimensional Higgs superfield $\hat{\phi}$. 
The superpotential includes the term
\begin{eqnarray*}
\hat{f} \ni f\hat{\psi}\hat{\psi}\hat{\phi} +\cdots
\end{eqnarray*}
responsible for quark and lepton masses, with $f$ the single Yukawa
coupling per generation in the $GUT$ scale theory.  The dots represent
terms including for instance higher dimensional Higgs representations
and interactions responsible for the breaking of $SO(10)$.  We neglect
Yukawa couplings and fermion masses for the first two generations. As
usual, SUSY breaking, is parametrized by associated soft SUSY breaking
mass terms. We assume a common mass $m_{16}$ for all matter scalars and
a mass
$m_{10}$ for the Higgs scalars, along with a universal gaugino mass
$m_{1/2}$, and common trilinear and bilinear SSB masses $A_0$ and
$B$. Motivated by apparent gauge coupling unification in the MSSM,
$SO(10)$ is assumed to break directly to the gauge group $SU(3)_C\times
SU(2)_L\times U(1)_Y$ at $Q=M_{GUT}$, where the reduction in rank of the
gauge symmetry leads to distinct $D$-term contributions to scalar field
mass terms. The scalar field squared masses at $Q=M_{GUT}$ are then
given by
\begin{eqnarray*}
m_Q^2=m_E^2=m_U^2=m_{16}^2+M_D^2 \\
m_D^2=m_L^2=m_{16}^2-3M_D^2 \\
m_{H_{u,d}}^2=m_{10}^2\mp 2M_D^2 , \\
\end{eqnarray*}
where $M_D^2$ parametrizes the magnitude of the $D$-terms,
and can, owing to our ignorance of the gauge symmetry breaking mechanism,
be taken as a free parameter, with either positive or negative values.
$|M_D|$ is expected to be of order the weak scale. 
Thus, the model is characterized by the following free 
parameters\footnote{If the effective theory below the GUT scale is the
MSSM plus a right-handed neutrino, then several other parameters
enter the discussion. Typically, these have only a small effect on the
SUSY particle mass spectrum: see Ref. \cite{bdqt}}:
\begin{eqnarray*}
m_{16},\ m_{10},\ M_D^2,\ m_{1/2},\ A_0,\ sign(\mu ).
\end{eqnarray*}
The value of $\tan\beta$ will be restricted by the requirement of
Yukawa coupling unification, and so is tightly constrained to a narrow
range around $\tan\beta\sim 50$.

In our previous paper\cite{bdft}, our procedure was as follows. 
We generated random samples of model
parameter sets within the ranges, 
\begin{eqnarray}
0&<&m_{16}<1500\ {\rm GeV},\nonumber\\
0&<&m_{10}<1500\ {\rm GeV},\nonumber\\
0&<&m_{1/2}<500\ {\rm GeV},\nonumber\\
-500^2\ {\rm GeV}^2 &<&M_D^2<+500^2\ {\rm GeV}^2,\label{rangesI}\\
45&<&\tan\beta <55, \nonumber\\
-3000\ {\rm GeV} &<& A_0<3000\ {\rm GeV}\ {\rm and} \nonumber\\
\mu &>&0\ {\rm or}\ \mu <0 .\nonumber
\end{eqnarray}
For each parameter set, we then calculated the non-universal scalar masses according to
formulae given above, and enter the parameters into the 
computer program ISASUGRA. 
ISASUGRA is a part of the ISAJET package\cite{isajet}
which calculates an iterative solution to the 26 coupled 
RGEs of the MSSM. 

To calculate the values of the Yukawa
couplings at scale $Q=M_Z$, we began with the pole masses $m_b=4.9$ GeV
and $m_\tau =1.78$ GeV. 
We calculated the corresponding running masses
in the $\overline{MS}$ scheme, and evolve $m_b$ and $m_\tau$ up to 
$M_Z$ using 1-loop SM RGEs. 
At $Q=M_Z$, we included the SUSY loop corrections to $m_b$ and
$m_\tau$ using the approximate formulae of Pierce {\it et
al.}\cite{pierce}, and then obtained the corresponding Yukawa couplings.
A similar procedure is used to calculate the top quark Yukawa coupling 
at scale $Q=m_t$. We assume a pole mass $m_t=175$ GeV. 

Starting
with the three gauge couplings and $t$, $b$ and $\tau$ Yukawa couplings 
of the MSSM at scale $Q=M_Z$ (or $m_t$), 
ISASUGRA evolves the various couplings up 
in energy until the scale where $g_1=g_2$, 
which is identified as $M_{GUT}$, is reached. The $GUT$
scale boundary conditions are imposed, and the full set of 26 RGE's
for gauge couplings, Yukawa couplings and relevant SSB masses are evolved
down to $Q\sim M_{weak}$, where the renormalization group improved
one-loop effective potential is minimized at an optimized scale choice
$Q=\sqrt{m_{\tst_L}m_{\tst_R}}$ and radiative electroweak symmetry breaking
is imposed. Using the new spectrum, the full set of SSB masses and couplings 
are evolved back up
to $M_{GUT}$ including weak scale sparticle threshold corrections to
gauge couplings. The process is repeated iteratively until a stable solution 
within tolerances is achieved. We accepted only solutions for which the 
Yukawa couplings $f_t$, $f_b$ and $f_\tau$ unify to 
within 5\%. This constraint effectively fixes the value of 
$\tan\beta$ to a narrow range of $49\pm 3$.
Yukawa unified solutions are found only for values of $\mu <0$.
This latter result is in accord with findings of 
Pierce {\it et al.}\cite{pierce} where solutions for $b-\tau$ unification 
in models with universality were found only for one sign of 
$\mu$ when $\tan\beta$ was large.
We also require the lightest
SUSY particle to be the lightest neutralino, and that electroweak symmetry 
is successfully broken radiatively. 

In this paper, we have used upgraded 
ISAJET 7.47 including 2-loop Yukawa coupling RGEs\cite{vb} with weak scale 
threshold corrections\cite{diego} to check our previous solutions.
The updated solutions as a function of model parameter space are 
presented in Fig. \ref{jav_new_1}; no apparent differences can be seen
compared with our earlier results which used just 1-loop RGEs for
Yukawa coupling evolution.
The Yukawa unified solutions to the SUSY particle mass spectra are found
to occur typically with $m_{16}\alt m_{10}\alt 1.5 m_{16}$
and for $0.1 m_{16}\alt M_D\alt .35 m_{16}$. Too small of a $D$-term will not
yield enough splitting in the Higgs boson soft masses to yield REWSB,
while too large a $D$-term will lead to third generation scalar field 
masses below experimental limits, or to charge and/or color breaking 
minima in the scalar potential. 
Our requirement of 5\% unification is determined by defining the 
variables $r_{b\tau}$, $r_{tb}$ and $r_{t\tau}$, where for instance
$r_{b\tau}= max(f_b/f_\tau,f_\tau /f_b )$. We then require
$R= max(r_{b\tau},r_{tb},r_{t\tau}) <1.05$. 

The methodology employed in Ref.~\cite{bdft} is useful for determining
the general regions of model parameter space where Yukawa-unified
solutions with REWSB can be found. However, in order to survey
the predictions of model parameter space for observable quantities
of experimental interest, it is useful to develop a more
systematic approach to the model parameter space. For most of the results
in this paper, we will take $\tan\beta =50$ as an indicative
central value. We are also restricted to $\mu <0$. Guided by the 
results of Fig. \ref{jav_new_1}{\it a}, we can also adopt a central value
$m_{10}=1.25 m_{16}$, and from Fig. \ref{jav_new_1}{\it b}, 
$M_D={1\over 5} m_{16}$ or $M_D={1\over 3} m_{16}$. Then, most of the 
variation in parameter space comes from the values of $m_{16}$ and
$m_{1/2}$, for which, from Fig. \ref{jav_new_1}{\it c}, there is no
obvious correlation. 

We show in Fig. \ref{pam_ps} the resulting
$m_{16}\ vs.\ m_{1/2}$ plane, taking as well $A_0=0$, for {\it a})
the smaller and {\it b}) the larger value of $M_D$. The region shaded 
by solid dots is excluded on the left by negative values of $m_A^2$,
and on the right by negative values of $\mu^2$: in either case, the 
REWSB constraint is violated. The unshaded region gives viable
solutions to the superparticle mass spectra, although Yukawa unification
can vary throughout the plot from regions with $R<1.05$ to $R\simeq 1.25$.
The value of $R$ can be fine-tuned to less than 1.05
throughout much of the parameter space plane by adjusting the 
value of $\tan\beta$ within the band in Fig.~\ref{jav_new_1}{\it d}. To gain an idea of the superparticle and Higgs boson
mass spectrum throughout the parameter space plane, we show dashed contours
of squark mass $m_{\tu_L}=1000$ (lower) and 2000 GeV (upper contour).
As noted in Ref. \cite{bdft}, the first and second generation squark masses
are bounded below by about 700 GeV, and are usually much heavier
in the Yukawa-unified $SO(10)$ model. The chargino mass contours of
$m_{\tw_1}=150$, 250 and 350 GeV are shown as solid contours, increasing
with $m_{1/2}$. They asymptote towards the right-hand parameter space
boundary, which is given by solutions where $\mu^2 =0$. For large 
values of the weak scale gaugino mass $M_2$, $|\mu |\alt M_2$, so that
the lightest chargino and neutralinos contain significant higgsino 
components. Although first and second generation squarks and sleptons
are quite heavy and beyond the reach of LEP2 and Tevatron experiments, 
the charginos can be light, and within the reach of these collider 
facilities. Finally, the dotted contours show values of the 
pseudoscalar Higgs mass $m_A= 150$, 250 and 350 GeV, from left to right.
Regions of parameter space with $m_A\simeq 100-200$ GeV and large 
$\tan\beta$, as in this model, may have Higgs bosons accessible
to Fermilab Tevatron collider searches via the $b\bar{b}A$ and
$b\bar{b}H$ production mechanisms\cite{run2higgs}.

In Fig. \ref{pam_a0}, we show similar plots of model parameter space,
this time taking $M_D={1\over 3} m_{16}$, but {\it a}) $A_0=-m_{16}$
and {\it b}) $A_0=+m_{16}$. The mass contours are the same as in 
Fig. \ref{pam_ps}. In Ref. \cite{bdft}, it was shown that the 
light bottom squark $\tb_1$ could well be by far the lightest of
all the squarks in Yukawa-unified $SO(10)$. For negative $A_0$
parameters, mixing effects in the bottom squark mass matrix
can give rise to even smaller values of $m_{\tb_1}\sim 100-300$ GeV, 
which may be accessible to direct searches at the 
Fermilab Tevatron\cite{bmt}. In addition, as can be seen, the range
of parameter space is increased beyond the corresponding result for 
$A_0=0$. Alternatively, taking $A_0=+m_{16}$ yields typically heavier 
bottom squark masses, and a smaller range of accessible parameter space.

In Fig. \ref{pam_yu}, we show again the parameter space plane for
$M_D={1\over 3} m_{16}$ but with {\it a}) $A_0=-m_{16}$ and {\it b})
$A_0=0$. The planes of Fig. \ref{pam_yu} shall form the template for
most of the results to come in the following sections. We show in
Fig. \ref{pam_yu} contours of the degree of Yukawa unification $R$ in
per cent. The Yukawa unification varied from less than 5\% to over 20\%
throughout the plane. We note that Fig. \ref{jav_new_1}{\it d}, makes it
evident that for lower values of $m_{1/2}$ we will have better
Yukawa coupling unification occurring at slightly lower values of
$\tan\beta$ than $\tan\beta=50$ in this figure.

It was pointed out in \cite{bdft} that the lightest slepton in this model 
is the stau, and that solutions with $m_{\tilde\tau_1}<200$ GeV are hard 
to obtain. This is confirmed in Fig.~\ref{st1ang_md} where we plot the stau
mass as a function of the left-right mixing angle in the stau sector. 
Four different regions are indicated according to the value of the D-term
mass marameter $M_D$ (there is some overlap between them in the boundaries).
As opposed to models with universality, where the light stau has mainly a
right stau component, in SO(10) it can have any value for the mixing 
angle. 
In ISAJET, we take $\ttau_1 =\cos\theta_\tau\ttau_L -\sin\theta_\tau
\ttau_R$, so that
small values of $M_D$ prefer $\cos\theta_{\tau}\approx0$ and a 
dominantly right $\ttau_1$, while large values of 
$M_D$ prefer $\cos\theta_{\tau}\approx1$ and a dominantly left $\ttau_1$.
Any intermediate value of $\cot\theta_\tau$ is also allowed, so that the
light stau could be a very mixed state of $\ttau_L$ and $\ttau_R$.

\section{Evaluation of neutralino relic density}

Recent precision measurements of cosmological parameters present new and
interesting constraints on the possibility of particle cold dark
matter\cite{turner}. Analyses of the cosmic microwave background
radiation suggest a matter/energy density of the universe $\Omega
={\rho\over\rho_c}=1.0\pm 0.2$, {\it i.e.} consistent with a flat
universe. In addition, the Hubble constant itself is now estimated as
$H_0=(65\pm 5)$ km/sec/Mpc, yielding a value $\Omega h^2\sim 0.42$,
where $H_0=100h$ km/sec/Mpc with $h$ the scaled Hubble constant. The
baryonic contribution to $\Omega h^2$ is estimated to be $\Omega_B
h^2\sim 0.02$ from Big Bang nucleosynthesis (BBN) arguments, while the
contribution from neutrinos ought to be $\Omega_\nu h^2\alt 0.06$,
mainly from models of structure formation in the universe.  Information
on the total matter density of the universe can be found by combining
constraints from BBN with measurements of intracluster gas and
gravitational binding in rich galactic clusters: the results indicate
$\Omega_M h^2\sim 0.17$.  The remaining matter density may come from
cold dark matter (CDM) particles, while the remaining energy density may
come from a non-zero cosmological constant, as is suggested by recent
measurements of type Ia supernovae.  The lightest neutralino of
supersymmetry is an excellent candidate for CDM in the universe.  In a
universe with 5\% each of baryonic matter and massive neutrinos, and a
cosmological constant with $\Omega_\Lambda h^2 \sim 0.25$, we would
expect $\Omega_{\tz_1} h^2 \sim 0.13$.  As an extreme, assuming no
contribution from the cosmological constant, we would obtain
$\Omega_{\tz_1} h^2 \sim 0.38$.In order to account only for the dark
matter needed to explain galactic rotation curves, we would expect
$\Omega_{\tz_1} h^2\agt 0.02$.

To estimate the relic density\cite{DNoj} of neutralinos in
Yukawa-unified $SO(10)$, we follow the calculational procedure outlined
in Ref. \cite{bb_relic}.  Briefly, we evaluate all tree level neutralino
annihilation diagrams exactly as helicity amplitudes. We then calculate
the neutralino annihilation cross section, and compute the thermally
averaged cross section times velocity using the fully relativistic
formulae of Gondolo and Gelmini\cite{gg}. Once the freeze-out
temperature is obtained via an iterative solution, we can
straightforwardly obtain the neutralino relic density
$\Omega_{\tz_1}h^2$. Our program takes special care to integrate
properly over any Breit-Wigner poles in $s$-channel annihilation
diagrams\cite{bb_relic}. We do not include $\tz_1-\ttau_1$
co-annihilation diagrams; these can be important in mSUGRA
models\cite{falk}, but are unimportant in the $SO(10)$ model since the
slepton masses are always far heavier than the lightest
neutralino.\footnote{Recently, two papers have appeared on calculating
the relic density in Yukawa unified $SO(10)$\cite{lazarides}. Since
these authors do not insist on correct third generation fermion masses,
they are able to obtain sparticle mass spectra without invoking
$D$-terms.  Thus, their results differ significantly from ours.} Our
calculations of the relic density also do not include effects that might
be significant~\cite{BDD} if the $b$-squark is sufficiently degenerate
with $\tz_1$.

Our results for the neutralino relic density are shown in
Fig. \ref{pam_rd} for the same parameters as in Fig. \ref{pam_yu}.  We
show contours of $\Omega h^2=0.02$, 0.1, 0.15, 0.2, 0.25, 0.3 and 0.35.
In frame {\it a}), we see that there exists wide ranges of parameter
space for which the relic density $0.02\alt \Omega_{\tz_1}h^2\alt 0.35$,
{\it i.e.} in the cosmologically interesting region. In no part of the
plane shown does the relic density exceed $\Omega_{\tz_1}h^2 =0.81$.
Much of the interesting region of parameter space occurs for large
values of scalar masses, which is contrary to the situation at low
$\tan\beta$, where large scalar masses suppress the annihilation cross
section, and typically yield $\Omega_{\tz_1}h^2 > 1$. At large
$\tan\beta$, the decay widths of the Higgs bosons $H$ and $A$ become
very large, typically $50-100$ GeV. Then annihilation can proceed through
the very broad $s$-channel $H$ and $A$ resonances, even if $2m_{\tz_1}\ne
m_A$ or $m_H$.  In a diagonal swath that splits the allowed parameter
space, the relic density drop to $\Omega_{\tz_1}h^2< 0.02$. This trough
is due to regions where $m_A\simeq 2m_{\tz_1}$, and where annihilation
can occur very efficiently through the $s$-channel diagrams.  In frame
{\it b}), we show similar results except for $A_0=0$.  Again, we see a
region between the $\Omega_{\tz_1}h^2 = 0.02$ with very low relic
densities. Throughout the remaining parameter space region, the relic
density is of cosmologically interesting values, and in fact never
exceeds $\Omega_{\tz_1}h^2\sim 0.6$.

\section{Direct detection rates for relic neutralinos}

A consequence of the SUSY dark matter hypothesis is that a
non-relativistic gas of neutralinos fills all space. To test this
hypothesis, a number of direct neutralino detection experiments have
been built or are under construction\cite{ddexp}. The general idea
behind these experiments is that neutralinos (or some other CDM
candidate) could elastically scatter off nuclei in some material,
transferring typically tens of keV of energy to the recoiling
nucleus. Current generation detectors are aiming at a sensitivity of 0.1-0.01
events/kg/day

The first step involved in a neutralino-nucleus scattering calculation is to
calculate the effective neutralino-quark and neutralino-gluon interactions.
The neutralino-quark axial vector interaction leads in the non-relativistic
limit to a neutralino-nucleon spin-spin interaction, which involves the
measured quark spin content of the nucleon. To obtain the neutralino-nucleus
scattering cross section, a convolution with nuclear spin form factors
must be made. The neutralino-quark and neutralino-gluon interactions
(via loop diagrams) can also resolve into scalar and tensor components.
These interactions can be converted into an effective scalar
neutralino-nucleon interaction involving quark and gluon parton distribution
functions. A neutralino-nucleus scattering cross section can be obtained
by convoluting with suitable scalar nuclear form factors. The final neutralino
detection rate is obtained by multiplying by the {\it local} neutralino
density (for which estimates are obtained from galaxy formation modeling), and
appropriate functions involving the velocity distribution of relic
neutralinos and the earth's velocity around the galactic center.

In this section, we present event rates for
direct detection of relic
neutralinos left over from the Big Bang. For illustration, we present detailed
calculations for a $^{73}$Ge detector, which has
a sizable nuclear spin content $J={9\over 2}$. We follow the 
procedures outlined in Ref. \cite{bb_dd}, and assume a
local neutralino density 
$\rho_{\widetilde{Z_1}}=5\times 10^{-25}\, \rm g\, cm^{-3}$\cite{flor}.

Our results are shown in Fig. \ref{pam_dd} for the same parameter
plane as in Fig. \ref{pam_yu}.
We show direct detection rate contours of 1, 0.1, 0.01 and 0.001 events/kg/day.
In the low $m_{1/2}$ regions of parameter space, the direct detection
rate exceeds 1 event/kg/day, which is a very large rate for neutralino 
dark matter. The large rate occurs generically at large 
$\tan\beta$\cite{dn,bb_dd}. Neutralinos can scatter off quarks via
squark or Higgs boson exchange graphs, and off gluons via loop
graphs containing quarks, squarks and gluons. Neutralino-gluon
scattering via Higgs exchange into a $b$-quark loop for instance is enhanced
at large $\tan\beta$, and helps lead to the large scattering rate.
Unfortunately, the largest direct detection rates occur mainly in those
regions of parameter space where the relic density is tiny.
However, there do exist significant regions of parameter space
where the direct neutralino detection rate exceeds 0.01 events/kg/day
where the relic density is comfortably within the cosmologically
interesting range.

\section{Constraint from radiative $B$ decay}

\subsection{Results for $BR(b\to s\gamma )$}

The radiative flavor-changing decay $b\to s\gamma$ can occur in the
SM via $Wt$ loop diagrams; in SUSY theories, additional diagrams
including $tH^+$, $\tw_i\tst_j$, $\tg\tq$ and $\tz_i\tq$ loops will
occur\cite{bbmr,morebsg}. The ALEPH collaboration\cite{aleph} has measured
\begin{equation}
BR(b\to s\gamma )=(3.11\pm 0.80\pm 0.72)\times 10^{-4}
\end{equation}
while the CLEO collaboration\cite{cleo} has recently reported
\begin{equation}
BR(b\to s\gamma )=(3.15\pm 0.35\pm 0.32\pm 0.26)\times 10^{-4}
\end{equation}
and restricts the branching ratio at 95\% CL to be
\begin{equation}
2\times 10^{-4}<BR(b\to s\gamma )<4.5\times 10^{-4} .
\end{equation}
The SM prediction at NLL accuracy\cite{bsg_sm} is
\begin{equation}
BR(b\to s\gamma )=(3.28\pm 0.33)\times 10^{-4} .
\end{equation}

The calculation of the width for $b\to s\gamma$ decay proceeds by
calculating the loop interaction for $b\to s\gamma$ within a given model
framework, {\it e.g.} the MSSM,
at some high mass scale $Q\sim M_W$, and then matching
to an effective theory Hamiltonian given by
\begin{equation}
H_{eff}=-{4G_F\over \sqrt{2}} V_{tb}V^*_{ts}\sum_{i=1}^8 C_i(Q )O_i(Q ),
\label{eq1}
\end{equation}
where the $C_i(Q )$ are Wilson coefficients evaluated at scale $Q$,
and the $O_i$ are a complete set of operators relevant for the process
$b \to s\gamma$ (given, for example, in Ref. \cite{gsw}.)
All orders approximate QCD corrections are included via renormalization group
resummation of leading logs (LL) which arise
due to a disparity
between the scale at which new physics enters the $b\rightarrow s\gamma$ loop
corrections (usually taken to be $Q\sim M_W$), and the scale at which
the $b\rightarrow s\gamma$ decay rate is evaluated ($Q\sim m_b$).
Resummation then occurs when we solve the renormalization
group equations (RGE's) for the Wilson coefficients
\begin{equation}
Q {d\over dQ} C_i(Q )=\gamma_{ji} C_j(Q ),
\label{eq2}
\end{equation}
where $\gamma$ is the $8\times 8$ anomalous dimension matrix (ADM),
and
\begin{equation}
\gamma={\alpha_s\over 4\pi}\gamma^{(0)}+({\alpha_s\over 4\pi})^2
\gamma^{(1)}+\ldots .
\label{eq3}
\end{equation}
The matrix elements of the operators
$O_i$ are finally calculated at a scale $Q\sim m_b$ and multiplied by
the appropriately evolved Wilson coefficients to obtain the final
decay amplitude.

Recently, next-to-leading order QCD corrections
have been completed for $b\to s\gamma$ decay. These include {\it i})
complete virtual corrections\cite{ghw} to the relevant
operators $O_2,\ O_7$ and $O_8$ which, when combined with
bremsstrahlung corrections\cite{brem,ghw}
results in cancellation of associated soft and collinear singularities;
{\it ii}) calculation of ${\cal O}(\alpha_s^2)$ contributions to
the ADM elements $\gamma_{ij}^{(1)}$
for $i,j=1-6$ (by Ciuchini {\it et al.}\cite{ciuchini}), for
$i,j=7,8$ by Misiak and M\"unz\cite{misiak1}, and for $\gamma_{27}^{(1)}$
by Chetyrkin, Misiak and M\"unz\cite{misiak2}. In addition, if
two significantly different masses
contribute to the loop amplitude, then there can already exist significant
corrections to the Wilson coefficients at scale $M_W$.
In this case, the procedure is to
create a tower of effective theories with which
to correctly implement the RG running
between the multiple scales involved in the problem.
The relevant operator bases, Wilson coefficients and
RGEs are given by Cho and Grinstein\cite{cg} for the SM and by
Anlauf\cite{anlauf} for the MSSM. The latter analysis includes contributions
from just the $tW$, $tH^-$ and $\tst_i\tw_j$ loops
(which are the most important ones). We include the above set of QCD
improvements (with the exception of $\gamma_{27}^{(1)}$, which has been shown
to be small\cite{misiak2})
into our calculations of the $b\to s\gamma$ decay rate for the mSUGRA model.

We include the contributions to $C_7(M_W)$ and $C_8(M_W)$ 
from $\tg\tq$ and $\tz_i\tq$ loops.
To do so, the squark mixing matrix $\Gamma$ which enters the couplings
must be derived. To accomplish this, 
we first calculate the values of all running fermion masses in the SM
at the mass scale $M_Z$. From these, we derive the corresponding
Yukawa couplings $f_u$, $f_d$ and $h_e$ for each generation, and construct
the corresponding Yukawa matrices $(f_u)_{ij}$, $(f_d)_{ij}$ and $(f_e)_{ij}$,
where $i,j=1,2,3$ runs over the 3 generations. We choose a basis that
yields flavor
diagonal matrices for $(f_d)_{ij}$ and $(f_e)_{ij}$, whereas the CKM mixing
matrix creates a non-diagonal matrix $(f_u)_{ij}$\cite{diego}.
The three Yukawa matrices are evolved within the MSSM from $Q=M_Z$
up to $Q=M_{GUT}$ and the values are stored. 
At $Q=M_{GUT}$, the matrices $(A_uf_u)_{ij}$, $(A_df_d)_{ij}$ and
$(A_ef_e)_{ij}$ are constructed (assuming $A(M_{GUT})=A_0 \times {\bf 1}$). The
squark and slepton mass squared matrices $(m^2_k)_{ij}$ are also
constructed, where
$k=\tilde Q, \tilde u, \tilde d, \tilde L $ and $\tilde e$. These
matrices are assumed to be proportional to {\bf 1} at $Q=M_{GUT}$.
The $(Af)_{ij}$ and $(m^2_k)_{ij}$ matrices are evolved along with
the rest of the gauge/Yukawa couplings and soft SUSY breaking terms
between $M_Z$ and $M_{GUT}$ iteratively via Runge-Kutta method
until a stable solution is found. 
At $Q=M_Z$, the $6\times 6$ $d$-squark mass squared
matrix is constructed.
Numerical diagonalization of this matrix yields the squark mass mixing matrix
$\Gamma$ which is needed for computation of the $\tg\tq$ and $\tz_i\tq$
loop contributions.
At this point, the Wilson coefficients $C_7(M_W)$ and $C_8(M_W)$ can
be calculated and evolved to $Q\sim m_b$ as described above, so that the
$b\to s\gamma$ decay rate can be calculated\cite{bsg1,bsg2}.

As indicated above, the main non-SM contributions to 
$B(b\rightarrow s\gamma)$ are $tH^{\pm}$ and $\tilde t_i\tw^{\pm}_j$. 
There is some correlation between the $H^{\pm}$ and $\tilde t_1$
masses and the D-term mass parameter $M_D$, as we show in 
Fig.~\ref{chhst1_md}. Large values of the charged Higgs are guaranteed if 
$M_D>350$ GeV, which involve also large values of the top squark mass. This is 
the most convenient situation for satisfying the experimental requirements 
for $B(b\rightarrow s\gamma)$, because the charged Higgs contribution 
decreases. Nevertheless, in order to decrease sufficiently the chargino 
contribution, it 
is necessary to work in a restricted region of parameter space, 
as indicated below. In contrast to the correlation in Fig.~\ref{chhst1_md}, 
there is no obvious correlation between the charged Higgs mass and the mass 
of the chargino.

Our results for the same parameter space planes as in Fig. \ref{pam_yu}
are shown in Fig. \ref{pam_bsg}. 
Contours of $B(b\to s\gamma )$ are shown for 5, 6, 7, 8, 10 and 
$15\times 10^{-4}$.
In frame {\it a}), for $A_0=-m_{16}$, we see that very large values of
$B(b\to s\gamma )$ are obtained, usually well above the CLEO 95\%
CL limit, except for the very largest values of $m_{1/2}\sim 1000$ GeV.
In particular, the regions with the maximum rate for
direct detection of dark matter also have $B(b\to s\gamma )$ values far
in excess of experimental measurements. A similar behavior is seen in frame
{\it b}) for $A_0=0$. The very large values of $B(b\to s\gamma )$
are typical of SUSY models with large values of $\tan\beta $ and $\mu <0$.
The qualitative behavior that Yukawa unification works best for
the sign of $\mu$ which gives the largest values for $B(b\to s\gamma )$
has been noted previously by a number of authors\cite{cw,bop}.

We do note here that there may be other possible effects which could act
to reduce the $b\to s\gamma $ branching fraction.  Nontrivial flavor
mixing in the down squark sector could result in significant
non-cancelling contributions from gluino loop diagrams, thus reducing
the value of $B(b\to s\gamma )$.  Furthermore, additional sources of
non-universality could act to yield values of $B(b\to s\gamma )$ more
nearly in line with experimental values\cite{bop}.  For instance,
non-universality between generations is still allowed by $SO(10)$, and
could lead to potentially large gluino loop contributions. This could,
for instance, naturally be the case in a class of $SO(10)$ models where
dynamics makes third generation sfermion masses much smaller than masses
of other sfermions~\cite{bmtimh}. Also, modifications can occur by
including large $CP$ violating phases for terms such as the $\mu$
parameter which controls the alignment of the $SM$ Wilson coefficients
with the top squark contribution.  Of course, such phases would also
affect for instance the electron and neutron electric dipole moments,
and even scattering cross sections\cite{brhlik_kane}. Finally, radiative
corrections due to Yukawa couplings are not included in the computation
of the decay rate. Keeping these issues in mind, one should exercise
judgement in drawing broad conclusions from the
results of Fig. \ref{pam_bsg}.

It is well known that the supersymmetric contribution
to $B(b\rightarrow s\gamma)$ is proportional to $A_t\mu\tan\beta$ when
$\tan\beta$ is large\cite{cmw}. Motivated by this, we have made a dedicated 
search in a restricted area of parameter space of our specific model
where $A_t$ is close to zero at the weak scale. 
To be more specific, we consider:
\begin{eqnarray}
1300\ {\rm GeV} <& m_{16} <& 1700\ {\rm GeV} \nonumber\\
1300\ {\rm GeV} <& m_{10} <& 1700\ {\rm GeV} \nonumber\\
 300\ {\rm GeV} <& M_D    <&  600\ {\rm GeV} \\
  50\ {\rm GeV} <& m_{1/2}<&  300\ {\rm GeV} \nonumber\\
 800\ {\rm GeV} <& A_0    <& 1800\ {\rm GeV} \nonumber
\end{eqnarray}
In Fig. \ref{mike_bsg}, we plot the branching ratio $B(b\rightarrow s\gamma)$ 
as a function of the light chargino mass, including only the points that 
satisfy the experimental constraints from colliders (aside from the
LEP2 chargino mass bound). 
We see that many solutions can occur with $B(b\to s\gamma )$ within
the 95\% CL range, and with a low value of $m_{1/2}$. 
These small $A_t$ solutions typically have $m_{10}<m_{16}$,
and very low chargino masses, but top squark masses above 1 TeV, which acts
to suppress $\tst_i\tw_i$ loop contributions. They 
correspond to solutions mostly to the left of the diagonal line
in the upper part of 
Fig. \ref{jav_new_1}{\it a}). 

\subsection{Comparison with SO(10) analysis of Blazek and Raby}

A recent study by Blazek and Raby has appeared\cite{br} (BR), 
which also focusses
on the rate for $b\to s \gamma$ decays in Yukawa unified $SO(10)$ models.
There are a number of differences between their study and ours.
The main difference is that BR, based on a global fit\cite{bcrw}
of model 4c of Anderson {\it et al.} (Ref~\cite{fmasses}), reject all models
with $\mu <0$ based on the generally large rate for $BR(b\to s \gamma )$
obtained. On the contrary, we reject all models with $\mu >0$ based on the
inability to achieve a high degree of Yukawa coupling unification.
BR adopt a top-down solution to renormalization group running, and
impose universality of all matter scalars at $Q=M_{GUT}$, although
they allow for arbitrary GUT scale values of $m_{H_u}$ and $m_{H_d}$,
whose values are fixed by imposing REWSB. This allows BR to generate 
solutions with very small values of $\mu$, so that their $\tz_1$
will be largely Higgsino-like. They also generate rather small 
SUSY loop corrections to $m_b$, ranging from $0\%-8\%$. These may be in 
part due to the small values of $\alpha_s(M_Z)$ they generate, ranging from
$0.11-0.18$. In contrast, for $\mu >0$, we generate SUSY loop
corrections to $m_b$ of order $10\%-40\%$ at large $\tan\beta$, in accord with
Pierce {\it et al.}, Ref. \cite{pierce}. 
Using a bottom-up approach to renormalization group running, 
the large SUSY loop corrections to 
fermion masses allow us to generate Yukawa unification only good to
$25\%-75\%$ for this sign of $\mu$ (again in accord with Pierce {\it et al.}). 
Thus, if we allow a much more liberal
constraint on Yukawa unification, we will also obtain $\mu >0$ solutions, 
which in general can give $BR(b\to s \gamma )$ rates more closely in accord
with measurements. We note that for $\mu >0$, we find REWSB can occur for
$\tan\beta \sim 50$, even assuming universality of {\it all} scalar 
masses (as mentioned earlier in Sec. I).

\subsection{Note added:}

After completion of this manuscript, a preprint 
appeared\cite{deboer} wherein it was claimed that NLO corrections to
the $b\to s\gamma$ loop diagrams\cite{cdgg,bmu} can lead to sign reversals
on certain amplitudes, especially at large $\tan\beta$. This effect leads
to rates for $b\to s\gamma$ decay in accord with theory predictions
for Yukawa unified models with $\mu <0$, while excluding models at large
$\tan\beta$ with $\mu >0$. 
In view of these (incomplete) calculations, it
appears to us that one should view the restrictions from one-loop calculations
of $b\to s\gamma$ decay with care. 
Rather than categorically excluding regions, it appears
to be best to wait until complete calciulations are available. It would
indeed be exciting if these corrections led to values of the 
$b\to s \gamma$ decay rate in agreement with experiment for the 
sign of mu favored by Yukawa unification.

\section{Prospects for Yukawa-unified $SO(10)$ at the Fermilab Tevatron}

In Yukawa unified $SO(10)$ models, as with most SUSY models, the light
Higgs boson is bounded by $m_h\alt 125$ GeV, and so may well be
accessible to Tevatron searches with high luminosity and improved
detectors\cite{conway,bht,cmw}. The $h$ boson will in general be
difficult to distinguish from a SM Higgs boson, so that signatures from
other sparticles or additional Higgs bosons would be useful.  In much of
the parameter space of Yukawa unified $SO(10)$ models, the gluinos,
squarks and sleptons are too heavy to result in observable signals for
the Fermilab Tevatron collider.  There are, however, three exceptions to
this. 1.) For low values of $m_{16}$, the pseudoscalar and heavy scalar
Higgs bosons $A$ and $H$ may be accessible to searches via $p\bar{p}\to
b\bar{b}A$ or $b\bar{b}H$. 2.) If $\tan\beta$ is on the lower end of
acceptable values ({\it e.g.} $\tan\beta\sim 46-49$), then some regions
of parameter space with low values of $m_{1/2}$ may be accessible to
trilepton searches.  3.) For large negative values of the $A_0$
parameter, the $\tb_1$-squark becomes light enough that there exists a
substantial production cross section for $\tb_1\bar{\tb}_1$
production. In this section, we address each of these
possibilities.\footnote{The effect of $SO(10)$ $D$-terms on SUSY cross
sections at the Fermilab Tevatron collider has been examined in
Ref. \cite{datta}. These authors do not insist on Yukawa unification, so
that any value of $\tan\beta$ is allowed.}

\subsection{Search for the $H$ and $A$ Higgs bosons}

Here, the search is for $b\bar{b}\phi$ events, where $\phi =A$ or $H$.
At large $\tan\beta \sim 50$ as in Yukawa unified $SO(10)$, 
the $b\bar{b}A$ and $b\bar{b}H$ couplings are enhanced, leading to large 
production cross sections.
For the Higgs mass range of interest, $H$ and $A$ typically decay to
$b\bar{b}$ with a $\sim 90\%$ branching fraction, so that the signal
will be four $b$-jet events, where two of the $b$s reconstruct
to the Higgs mass. For Higgs masses in the range of interest, 
usually $m_H\sim m_A$, so that the resonances from each will be overlapping.
The dominant backgrounds come from $b\bar{b}b\bar{b}$, $b\bar{b}jj$,
$Wb\bar{b}$ and $Zb\bar{b}$ production. 

In Fig. \ref{mam10_md} we plot allowed regions in the $m_{10}$-$m_A$ plane 
when the independent parameters of the theory are varied as indicated in 
Eq.~(\ref{rangesI}). Four regions are displayed according to the value of 
the D-term mass parameter $M_D$ (the four regions have a small overlap 
in their boundaries). The following correlation is obvious from the figure: a
smaller CP-odd Higgs mass $m_A$ is obtained for smaller $M_D$ and
$m_{10}$ mass parameters. We have checked that the heaviest CP-even Higgs
mass $m_H$ is very close to $m_A$, in fact they are within $\pm 2\%$ ($\pm 
5\%$) of each other for $m_A>200$ (100) GeV. On the contrary, if $m_A$ is 
decreased below 100 GeV, $H$ becomes heavier than $A$, such that $m_H$ can 
be up to $50\%$ larger than $m_A$ for $m_A\sim 80$.

For the recently concluded year-long Tevatron workshop on SUSY/Higgs 
physics at Run 2, analyses have been performed to estimate
the reach for $H$ and $A$ at large $\tan\beta$ for both the $D0$ and
$CDF$ detectors\cite{run2higgs}. In this report, details of 
expected event rates, backgrounds and selection cuts can be found.
For $\tan\beta =50$, the $5\sigma$ discovery reach of the $D0$ 
detector is expected to be
$m_A\simeq 145, $ 190 and 220 GeV for integrated luminosities of
2, 10 and 30 fb$^{-1}$, respectively. Similarly, the reach of the CDF detector
is estimated to be $m_A\simeq 130,$ 160 and 220 GeV for the 
same integrated luminosities.

In Fig. \ref{pam_ha}, we show contours of $m_A$ that correspond to the
$5\sigma$ reach of the $D0$ detector, for the same parameter choices as
in Fig. \ref{pam_yu}, assuming integrated luminosities of 2, 10 and 30
fb$^{-1}$.  The region to the left of the contours ought to be
accessible to Tevatron searches for $b\bar{b}H$ and $b\bar{b}A$
events. These regions represent the most likely possibility for
detection of beyond the SM physics at the Tevatron if Yukawa unified
$SO(10)$ is indeed correct.

\subsection{Search for trilepton events}

One of the most promising ways to detect supersymmetry at the Tevatron
for models with universality and low values of $\tan\beta$ is via
trilepton events, which dominantly come from 
$\tw_1\tz_2\to 3\ell +\eslt$\cite{trilep}. At large values of $\tan\beta$
and low values for scalar masses, the branching fraction of $\tz_2$ is 
dominated by real or virtual staus, so that the $\tz_2$ branching 
fraction to $e$s or $\mu$s is suppressed. Models with $m_{\tw_1}\alt 200$ GeV
have the potential to yield significant trilepton signals at the 
Fermilab Tevatron $p\bar{p}$ collider.

In a recent paper\cite{new3l}, 
the trilepton signal at the Fermilab Tevatron was 
re-examined for the mSUGRA model, in light of new backgrounds from
$W^*Z^*$ and $W^*\gamma^*$ production and decay, and the inclusion of
decay matrix elements into ISAJET. New sets of soft (SC2) and hard (HC2) 
cuts were proposed which allowed the trilepton signal to be seen at 
the Tevatron collider for integrated luminosities of 2 and 25 fb$^{-1}$.
For the Yukawa unified $SO(10)$ model, we have used the background rates 
and the SC2 cuts presented in Ref. \cite{new3l} to evaluate 
the observability of various points in model parameter space.

In scans of the parameter space corresponding to Fig. \ref{pam_yu}, 
no points were found where the Tevatron could see the $3\ell$ signal
using cuts SC2 with either 2 or 25 fb$^{-1}$ of integrated luminosity.
For both frames {\it a}) and {\it b}), 
the bottom squark $\tb_1$ is relatively light compared to other scalars
so that $\tz_2$ dominantly decays to $b\bar{b}\tz_1$ final states, 
at the expense of $e$ and $\mu$ modes.

By modifying the parameter space somewhat, it is possible to find
significant regions where the $3\ell$ signal should be visible.
In Fig. \ref{pam_3l}, we show the $m_{16}\ vs.\ m_{1/2}$ plane for
$m_{10}=1.14 m_{16}$, $A_0=0$, $\tan\beta =47$ and $M_D={1\over 3} m_{16}$.
The mass contours shown are the same as in Fig. \ref{pam_ps}.
In this case, there is a significant region below the $m_{\tw_1}=150$ GeV
contour accessible to Tevatron searches. 
No points were found to be observable at $5\sigma$ with 
just 2 fb$^{-1}$ of integrated luminosity, using cuts SC2. However, 
the open squares denote points where a 
$5\sigma$ $3\ell$ signal can be seen above background with 25 fb$^{-1}$.
The diamonds denote points seeable at $3\sigma$ level with 25 fb$^{-1}$
of data. In the region of parameter space with $m_{\tw_1}<150$ GeV
but with low values of $m_{16}$, the $\tb_1$ is becoming sufficiently 
light that three body $\tz_2$ decays to $b$-quarks dominates. 
In the visible region at higher values of $m_{16}$, the squarks, including
$\tb_1$ are so heavy that the $\tz_2$ decay is dominated by the 
virtual $Z$ contribution, and $\tz_2\to e^+e^-\tz_1$ with typically a
3\% branching fraction.
If we use the same parameters as in Fig. \ref{pam_3l} 
but increase $\tan\beta$ to 50, then 
the $\tb_1$ becomes so light that again $\tz_2$ dominantly decays to 
$b$-quarks, and the trilepton signal vanishes.

\subsection{Search for direct production of $b$-squarks}

In the Yukawa unified $SO(10)$ model, it has been noted in
Ref. \cite{bdft} that a combination of the large $b$-quark Yukawa
coupling plus the $D$-term contribution to the $\tb_R$ squared mass
yields parameter ranges with the light $b$-squark $\tb_1$ being by far
the lightest of all the squarks. In fact, frequently $\tb_1$ is also
much lighter than all the sleptons. This means that there may be regions
of model parameter space where $\tb_1\bar{\tb}_1$ pairs can be produced
at large rates, with possibly observable signals.

In Fig. \ref{sb1m16_md} we show the lightest bottom squark 
mass $m_{\tilde b_1}$ 
as a function of $m_{16}$ for different values of the D-term parameter $M_D$.
The three regions correspond to $M_D<220$~GeV, 220~GeV $<M_D<330$~GeV, and 
$330<M_D$~GeV, with some overlap at the boundaries. The smallest
bottom squark masses we obtain 
satisfy $m_{\tilde b_1}<200$ GeV and occur for
intermediate values of $M_D$ and $m_{16}$. These points have
approximately $\mu\sim-200$ GeV imposed by EWSB (and $A_0\sim-1600$ GeV).
This value of $\mu$ together with the large value of $\tan\beta$, makes
the left-right sbottom mixing comparable with the right sbottom soft mass
parameter $m_D$.

In Ref. \cite{bmt}, the reach of the Tevatron collider for $\tb_1\bar{\tb}_1$
pairs was investigated. If $\tb_1\to b\tz_1$ is the dominant decay mode,
then the final state will include two hard $b$-jets plus $\eslt$.
Backgrounds investigated included $W+jets$, $Z+jets$ and $t\bar{t}$
production. For the vector boson production backgrounds, the $jets$
qualifier includes both $b$-jets and other light quark jets. 
A number of cuts were proposed, including tagging the $b$-jets via
microvertex detectors, which allowed signals to be seen above backgrounds
in large parts of the $m_{\tb_1}\ vs.\ m_{\tz_1}$ parameter space.
Typically, bottom squark masses of $m_{\tb_1}\simeq 210$ GeV could be
seen with 2 fb$^{-1}$, and $m_{\tb_1}\simeq 240$ GeV could be seen
with 25 fb$^{-1}$. The reach was somewhat reduced if $\tb_1\to b\tz_2$
decays were also linematically accessible.

In Fig. \ref{pam_b}, we show the same parameter space as in 
Fig. \ref{pam_yu}{\it a}). We have generated events with ISAJET
for $p\bar{p}$ collisions at $\sqrt{s}=2$ TeV using the 
detector simulation, cuts and background rates described in Ref. \cite{bmt}.
Parameter space points labelled with a star are seeable at the $5\sigma$
level with just 2 fb$^{-1}$ of integrated luminosity. In addition, points
labelled with an open square are seeable at 25 fb$^{-1}$. 
The open squares have $\tb_1$ masses as heavy as
245 GeV.

\section{Summary, conclusions and outlook}

We have shown before that supersymmetric particle spectra can be calculated
in minimal supersymmetric $SO(10)$ GUT models consistent with Yukawa
coupling unification to 5\% or better, a top mass of $m_t=175$ GeV
and including radiative electroweak symmetry breaking. To accomplish this,
we made crucial use of $D$-term contributions to scalar masses that are 
induced by the reduction in rank of gauge symmetry when $SO(10)$ breaks
to the MSSM at $Q=M_{GUT}$. The $D$-terms split the GUT scale values of 
the Higgs boson squared masses so that $m_{H_u}^2<m_{H_d}^2$ already at 
$M_{GUT}$. This allows $m_{H_u}^2$ to be driven more negative than
$m_{H_d}^2$ at $Q=M_{weak}$, which is required for REWSB to take place.
We included the essential SUSY loop corrections to 
third generation fermion masses to generate the appropriate weak scale 
Yukawa couplings.
Yukawa unification to 5\% takes place only for $\mu <0$ and 
$\tan\beta\sim 46-52$.
In this paper, we upgraded our Yukawa coupling evolution equations to
include two-loop terms in the RGEs plus weak-scale threshold
effects for the one-loop terms.

The sparticle mass spectra we calculate reflect the influence
of the $D$-terms, and as a consequence, the $\tb_1$ turns out to be almost
always the lightest of the squarks, and is sometimes even lighter than
the sleptons. In addition, the lightest sleptons can frequently be 
(dominantly) left- sparticles, as opposed to the mSUGRA model where they
are usually right- sparticles. This latter trait may be tested explicitly at 
$e^+e^-$ linear colliders operating above slepton pair production threshold.

We attempted to systematize the parameter space of these models, showing
plots of sparticle mass in the $m_{16}\ vs.\ m_{1/2}$ plane for viable
selections of $m_{10}$, $M_D$ and $\tan\beta$. Of course, fixing these 
parameters meant we had to loosen our restrictions on the degree of
Yukawa coupling unification. We stress though that the original degree of
unification can be regained by a small change in $\tan\beta$. 

We calculated the relic density of neutralinos in Yukawa unified $SO(10)$,
and found none of the parameter space that was explored to be excluded
by the limits on the age of the universe. 
In fact, very desirable values for $\Omega_{\tz_1}h^2$ were found
throughout much of parameter space.

We also evaluated the direct neutralino detection rate for a $^{73}$Ge
detector. Detection rates could exceed 1 event/kg/day, but only in 
parameter regions with very low relic densities. In more favorable regions, 
the direct detection rate often exceeded 0.01 events/kg/day, which is the 
capability being sought by current generation dark matter 
detection experiments.

We also examined the branching ratio for $b\to s\gamma$ decay in Yukawa
unified $SO(10)$. We found that it was beyond the 95\%CL
experimental limits for 
much of the parameter space shown. However, special regions of parameter
space giving rise to small values of weak scale $A_t$ gave more acceptable
$B(b\to s\gamma )$ values, but with $m_{\tst_1} \geq 1$~TeV.
There could be other mechanisms
such as non-trivial mixing in the squark sector, additional
non-universality or mixing, or $CP$ violating phases
could act to reduce the large $B(b\to s\gamma )$ values.
Alternatively, the rate for $b\to s\gamma$ is more 
favorable for $\mu >0$, but then the degree of Yukawa unification would have
to be relaxed substantially. Finally, recent claims have been made\cite{deboer}
that two-loop $b\to s\gamma$ calculations can actually reverse the signs 
of certain decay amplitudes, leading to the possibility of agreement
between $b\to s\gamma$ calculations and Yukawa unified SUSY models
with $\mu <0$. In light of these claims, it seems prudent to refrain from
excluding parameter space regions based on one-loop $b\to s\gamma$ 
calculations until the theoretical predictions become more certain.

We found three possibilities for new particle detection at the 
upgraded Fermilab 
Tevatron beyond discovery of the lightest Higgs boson. In regions of
parameter space with small $m_{16}$, $b\bar{b}A$ and $b\bar{b}H$
production could be visible. In some parameter space regions with 
small $m_{1/2}$, 
it is possible to detect trilepton signals from $\tw_1\tz_2\to 3\ell$
production. In other regions again with small $m_{1/2}$, bottom squark
pair production may yield a visible signal. The other scalar particles
are generally too heavy to be produced substantially at the Tevatron
collider.

At the LHC, we expect Yukawa unified $SO(10)$ to be visible as usual for
values of $m_{\tg}$ beyond $\sim 2$~TeV. Within this framework, SUSY events
ought to be rich in $b$-quark jets due to the relatively light third
generation squark masses inherent in this theory.

Yukawa unified $SO(10)$ models have been historically very compelling models
for reasons listed in the introduction, and are even more compelling today
due to recent strong evidence on neutrino masses. The Yukawa
unified $SO(10)$ model is only one of a number of $SO(10)$ models that
are possible. A more general $SO(10)$ model\cite{bdqt} could contain
a superpotential with terms such as
\begin{equation}
\hat{f} = f_b\hat{\psi}\hat{\psi}\hat{\phi}_d+
f_t\hat{\psi}\hat{\psi}\hat{\phi}_u +\cdots
\end{equation}
in which case just $b-\tau$ unification would occur. This would
allow a much wider range of $\tan\beta\sim 30-50$ to occur\cite{bdqt}.

In addition, recent work on radiatively driven inverted hierarchy
models has shown that ``natural'' models can be obtained with multi-TeV
scale scalar masses, where the third generation masses are driven to 
weak scale values\cite{imh}. 
These models suffer the same problem with REWSB 
(and possible color and charge breaking minima in the scalar potential)
as the $SO(10)$ model considered here.
It has recently been shown that
$D$-term contributions to scalar masses aid this class of models to acheive
REWSB\cite{bmtimh}.
Finally, more complex gauge symmetry breaking patterns could occur which
would affect scalar masses in alternative ways\cite{kawamura}.

%
\acknowledgments
We thank Diego Casta\~no, Damien Pierce and Konstantin Matchev for discussions.
This research was supported in part by the U.~S. Department of Energy
under contract numbers DE-FG02-97ER41022 and DE-FG03-94ER40833. M.A.D. was supported in part by 
CONICYT-1000539.
%

%

\newpage
%
%

\iftightenlines\else\newpage\fi
\iftightenlines\global\firstfigfalse\fi
\def\dofig#1#2{\epsfxsize=#1\centerline{\epsfbox{#2}}}

%

%
\begin{figure}
\dofig{5in}{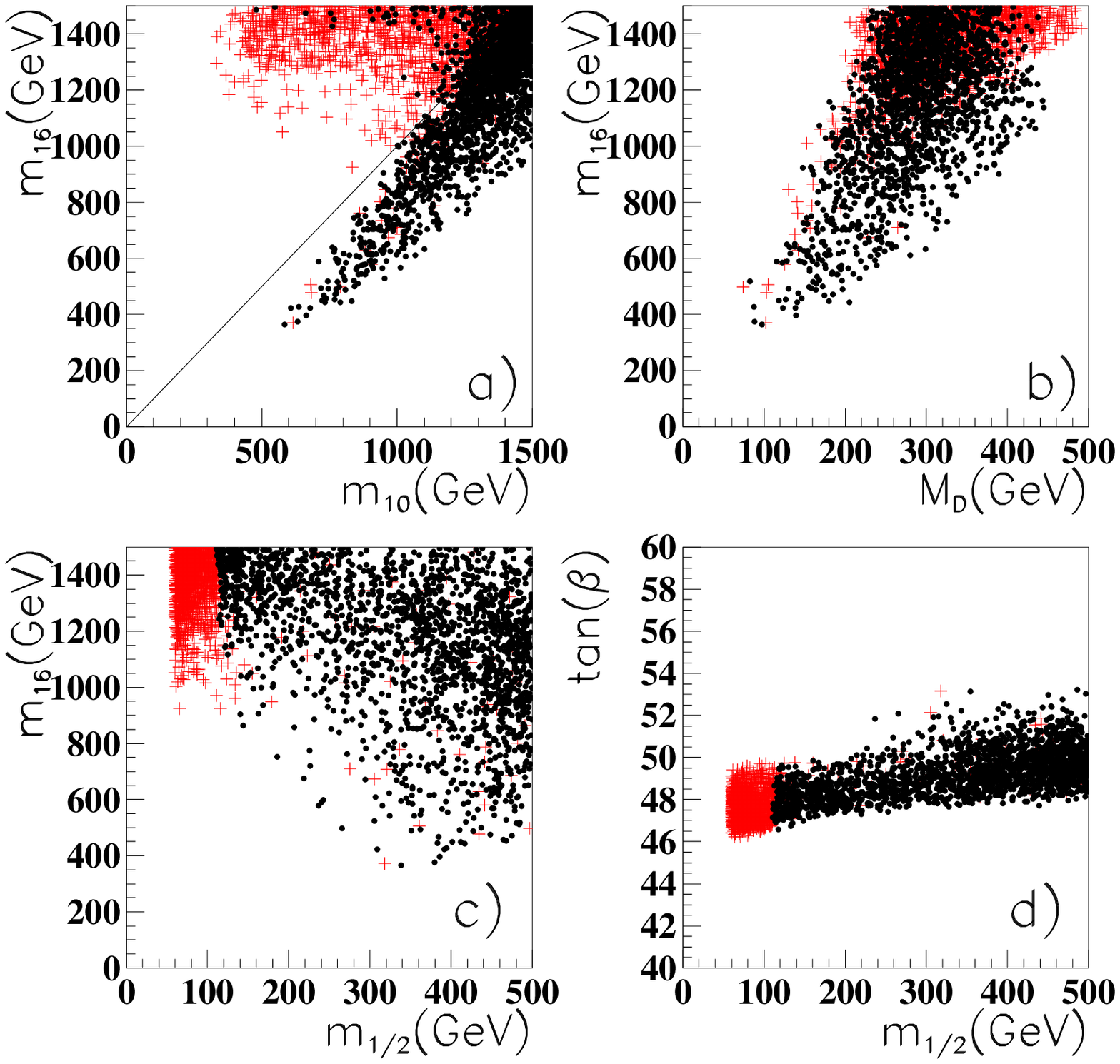}
\caption[]{
Plots of regions of parameter space where valid solutions to
minimal SUSY $SO(10)$ are obtained, consistent with Yukawa
coupling unification to 5\%, and radiative electroweak symmetry breaking.
We use 2-loop RGEs for both gauge and Yukawa coupling evolution.}
\label{jav_new_1}
\end{figure}
\begin{figure}
\dofig{7in}{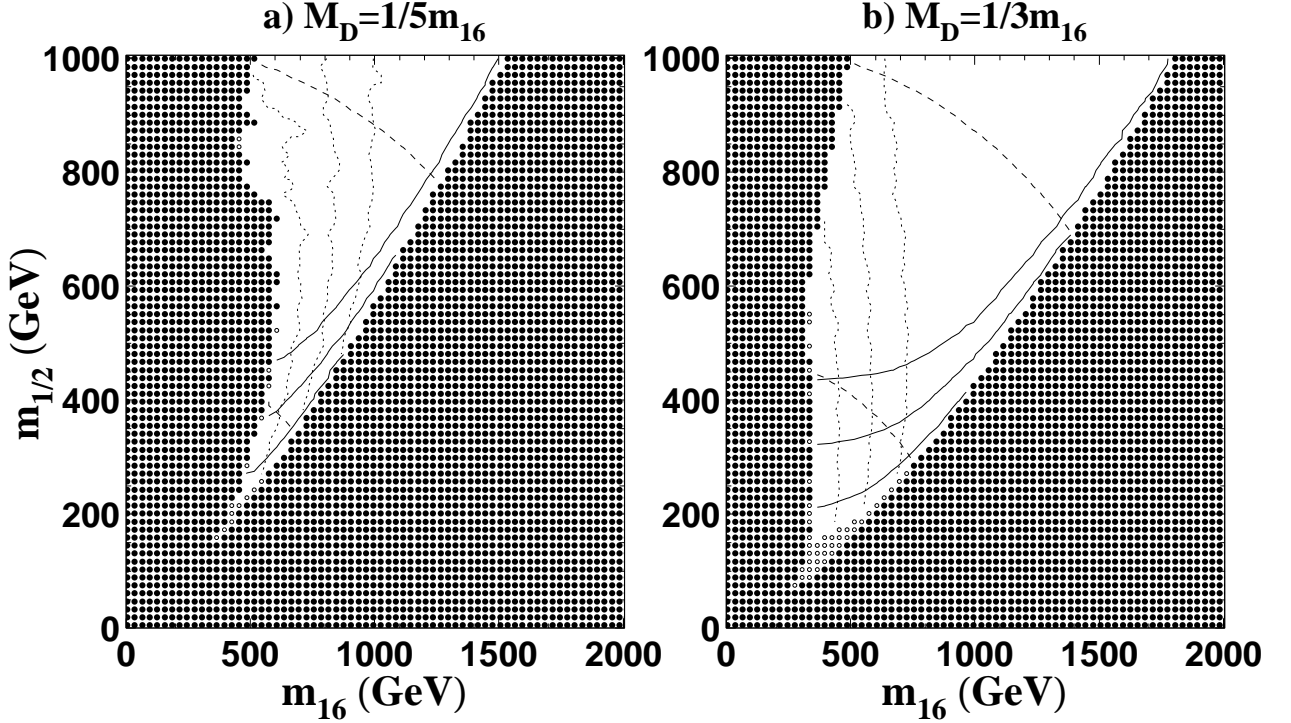}
\caption[]{
We plot theoretically allowed regions of $m_{16}\ vs.\ m_{1/2}$
parameter space taking $A_0=0$, $m_{10}=1.25m_{16}$, $\tan\beta =50$
and $\mu <0$. In frame {\it a}) we take $M_D={1\over 5}m_{16}$, while
frame {\it b} we take $M_D={1\over 3}m_{16}$. 
The shaded areas are not allowed by the REWSB constraint.
Open dots represent experimentally excluded points.
The dashed contours, from bottom to top, are for
$m_{\tu_L}=1000$ and 2000 GeV. The solid contours, from bottom to top,
represent $m_{\tw_1}$ masses of 150, 250 and 350 GeV. The dotted 
contours represent from left to right $m_A=150$, 250 and 350 GeV.
Yukawa coupling unification varies from less than 5\% in the central 
regions of the plot, to nearly 25\% at the edges of the excluded region.}
\label{pam_ps}
\end{figure}
\begin{figure}
\dofig{7in}{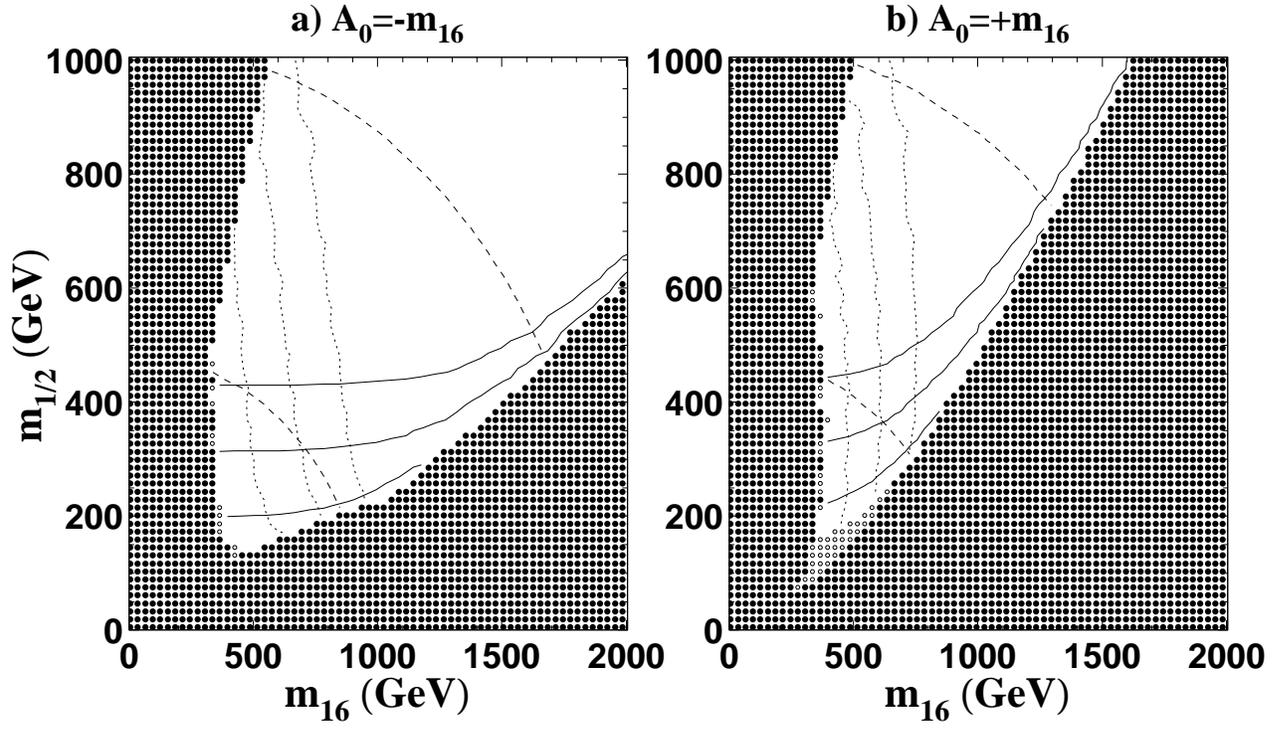}
\caption[]{
The same as Fig. \ref{pam_ps}, except $M_D={1\over 3}m_{16}$, while in
{\it a}), $A_0=-m_{16}$ and in {\it b}), $A_0=+m_{16}$.}
\label{pam_a0}
\end{figure}
\begin{figure}
\dofig{7in}{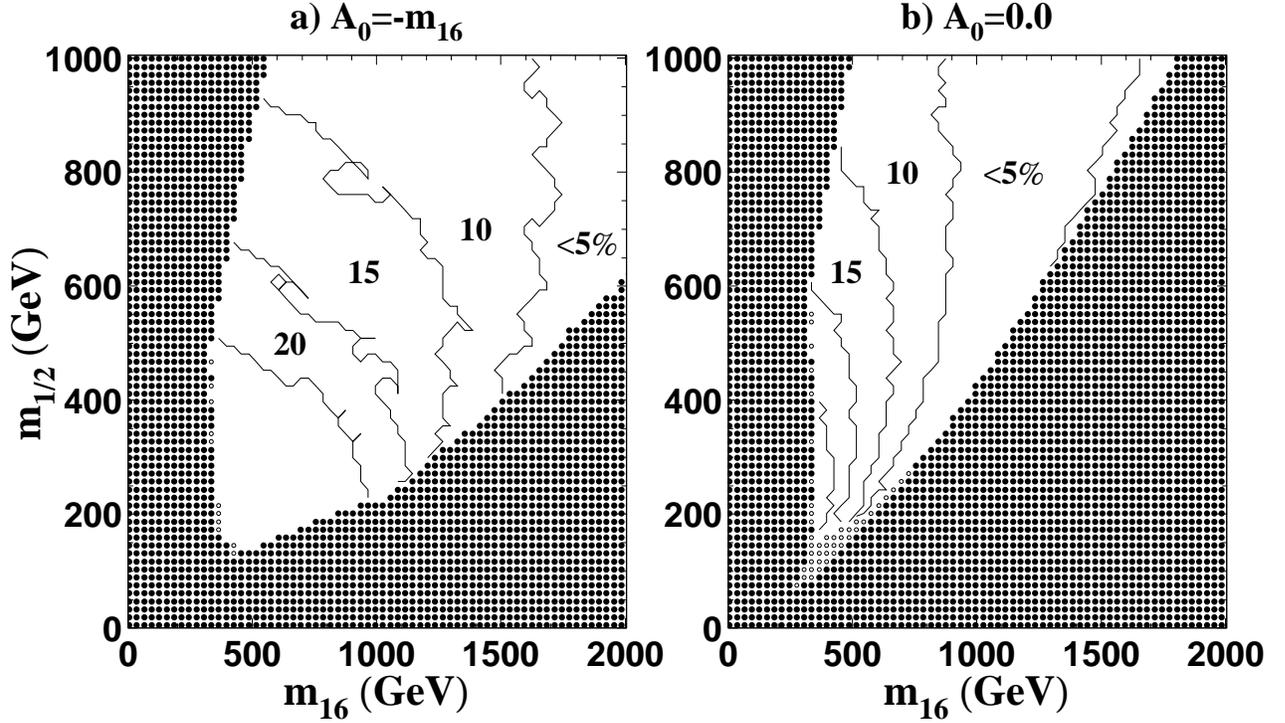}
\caption[]{
Contours showing Yukawa coupling unification in the $m_{16}\ vs.\ m_{1/2}$
plane, for $M_D={1\over 3}m_{16}$, $\tan\beta =50$, $m_{10}=1.25 m_{16}$,
$\mu <0$ and {\it a}) $A_0=-m_{16}$ and {\it b}) $A_0=0$. In frame {\it
b}) the degree of unification reduces to 5-10\% beyond the
right-most contour in the figure.}
\label{pam_yu}
\end{figure}
\begin{figure}
\dofig{6in}{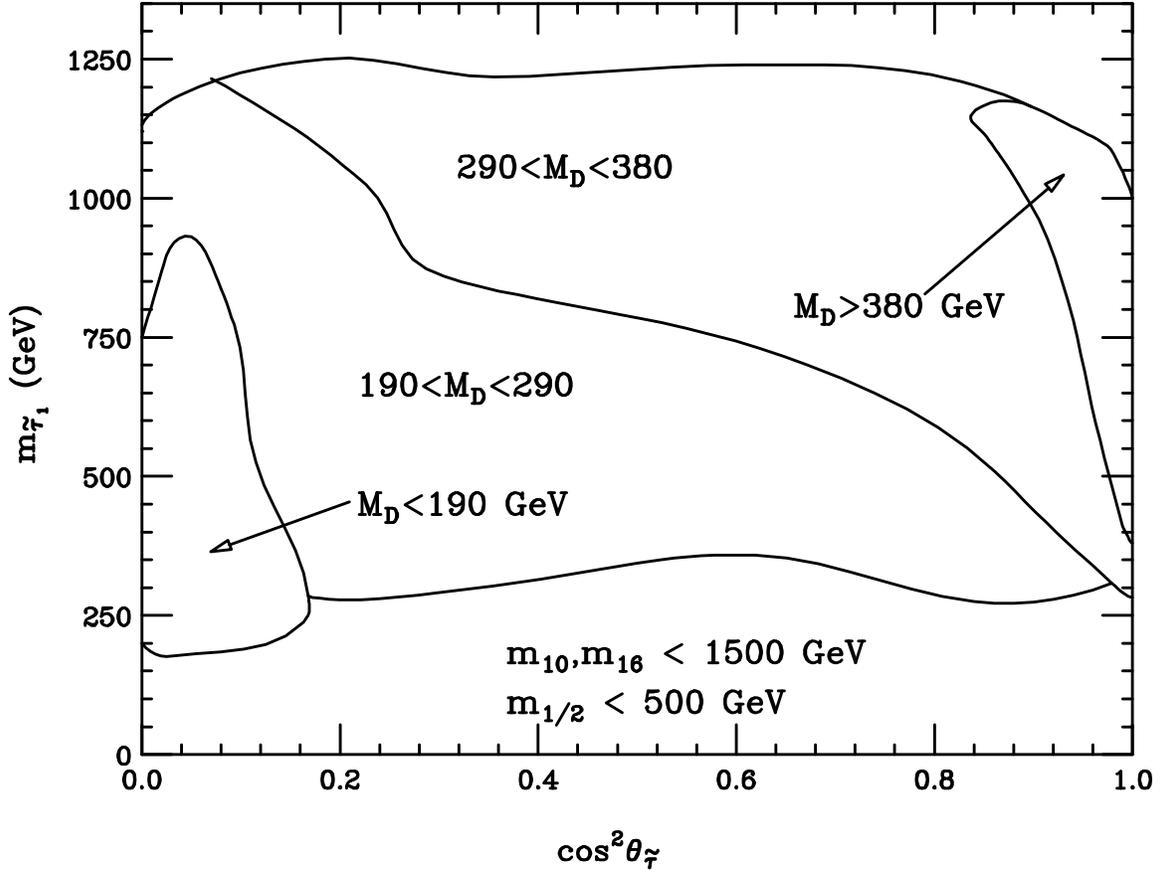}
\caption[]{Lightest stau mass as a function of the stau mixing parameter 
$\cos^2\theta_{\tau}$ for different values of the D-term parameter $M_D$.
In $SO(10)$ the stau mixing can take any value.}
\label{st1ang_md}
\end{figure}
\begin{figure}
\dofig{7in}{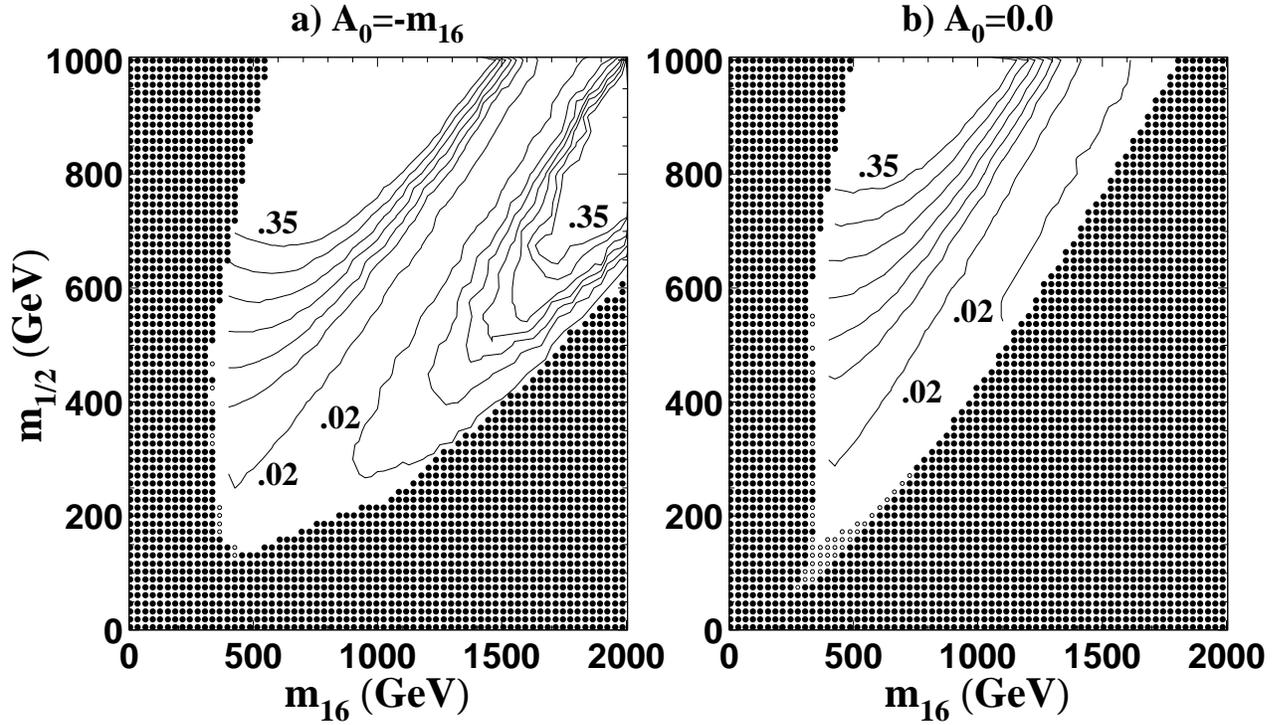}
\caption[]{
The same parameter space planes as in Fig. \ref{pam_yu}, but now showing 
contours of neutralino relic density 
$\Omega_{\tz_1} h^2=0.02$, 0.1, 0.15, 0.2, 
0.25, 0.30 and 0.35.}
\label{pam_rd}
\end{figure}
\begin{figure}
\dofig{7in}{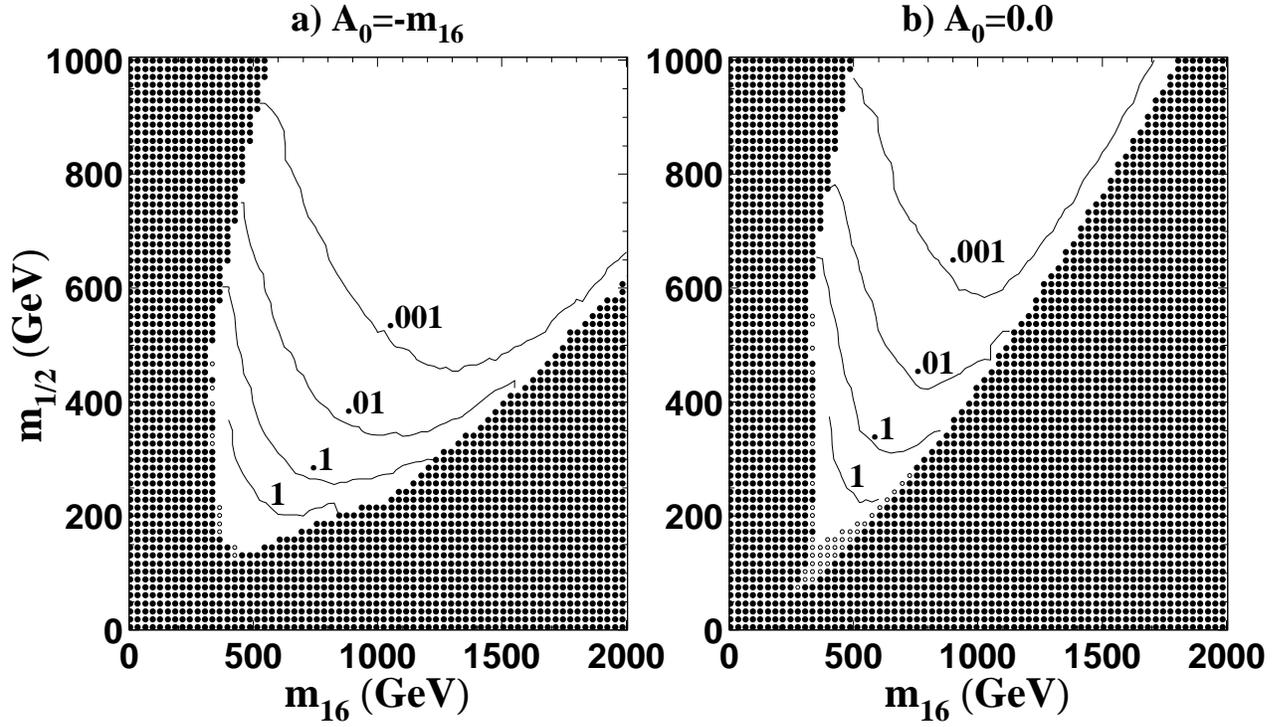}
\caption[]{
The same parameter space planes as in Fig. \ref{pam_yu}, but now showing 
contours of direct neutralino detection rate in events/kg/day in a 
$^{73}$Ge detector.}
\label{pam_dd}
\end{figure}
\begin{figure}
\dofig{6in}{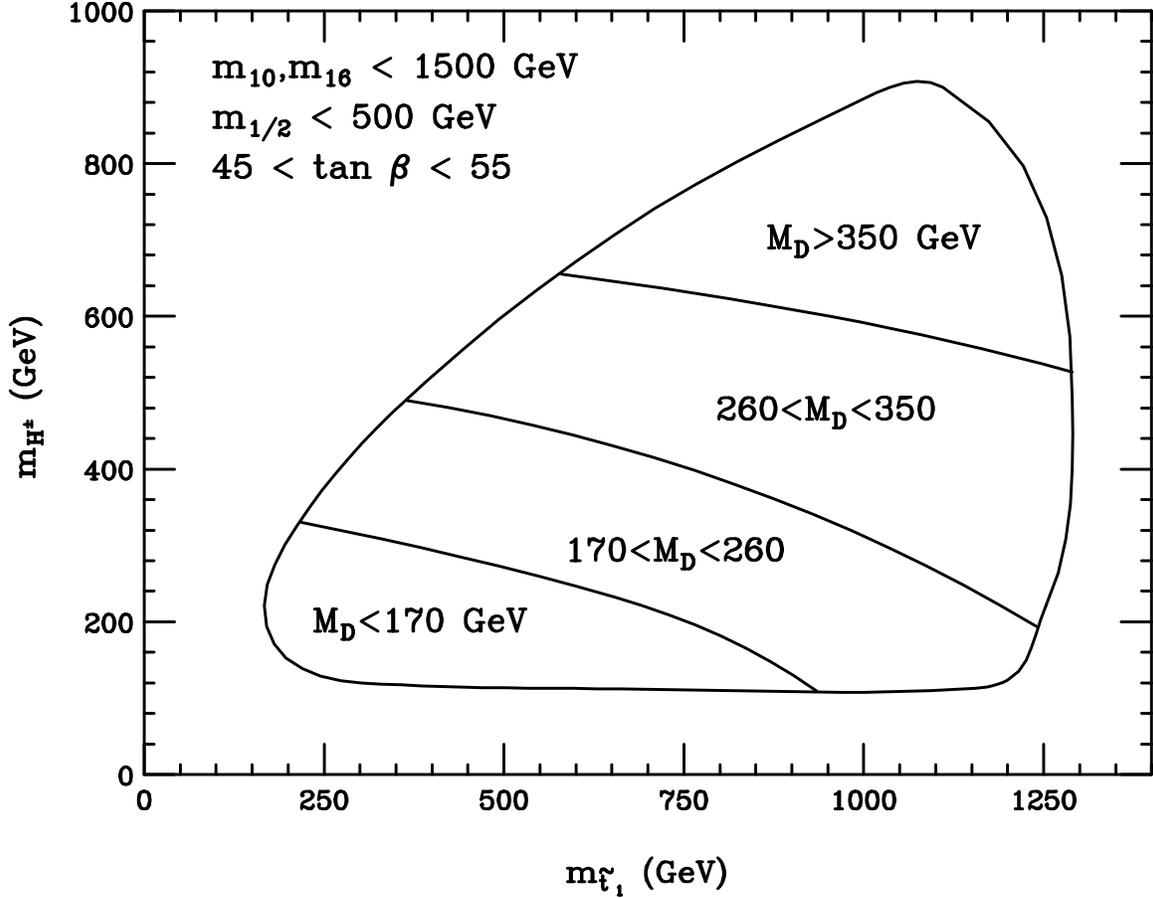}
\caption[]{Charged Higgs mass as a function of the light top squark mass for
different values of the D-term parameter $M_D$. Heavy $H^{\pm}$ and 
$\tilde t_1$ are associated with large values of $M_D$, which is favoured
by $B(b\rightarrow s\gamma)$. Here, $\mu <0$ and $|A_0| < 3$~TeV.}
\label{chhst1_md}
\end{figure}
\begin{figure}
\dofig{7in}{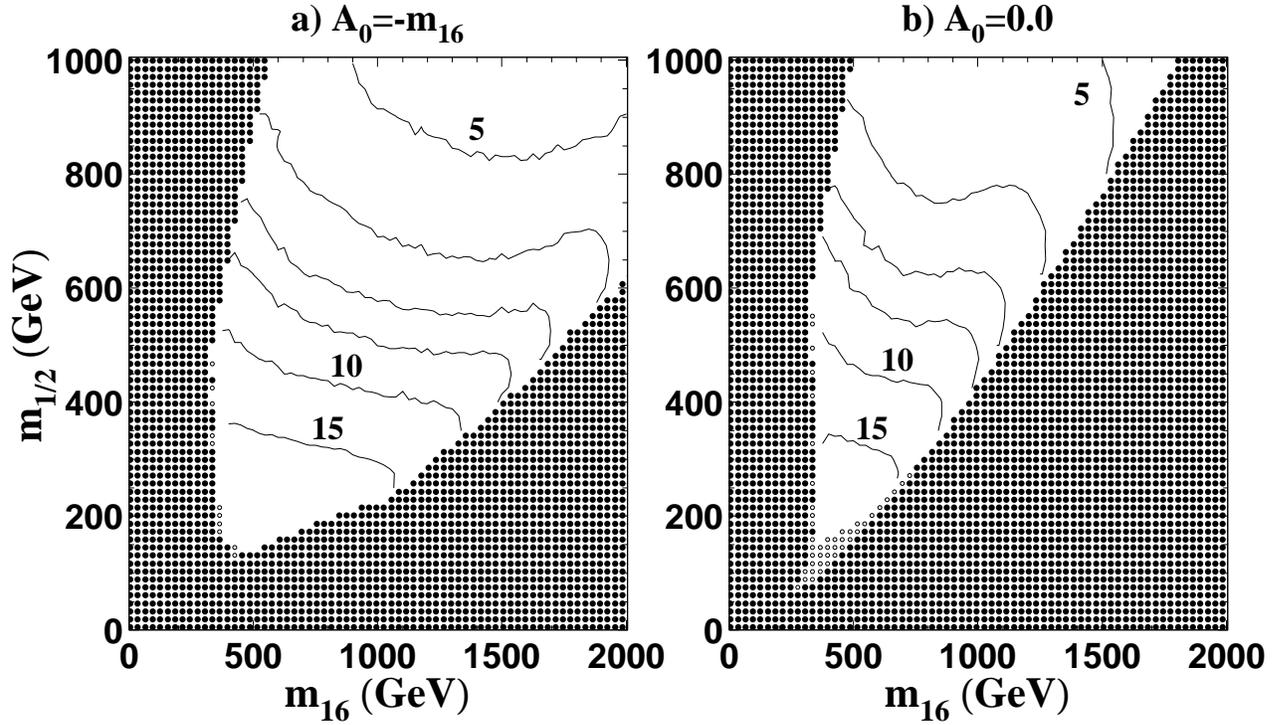}
\caption[]{
The same parameter space planes as in Fig. \ref{pam_yu}, but now showing 
contours of decay branching fraction for $b\to s\gamma$.
Starting from the top, the contours are for a branching fraction of
5,6,7,8,10 and 15 times $10^{-4}$.}
\label{pam_bsg}
\end{figure}
\begin{figure}
\dofig{6in}{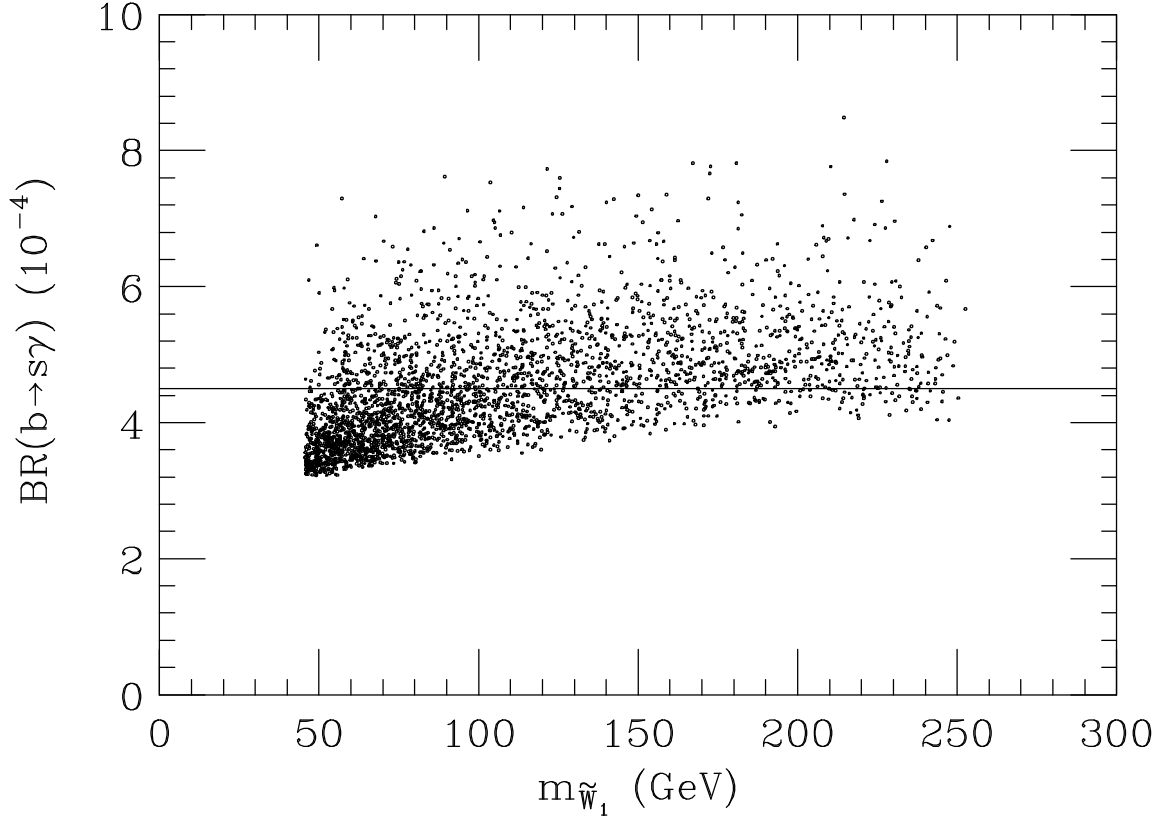}
\caption[]{Branching ratio $B(b\rightarrow s\gamma)$ as a function of the
light chargino mass when the input parameters are varied within the 
intervals given in Eq.~(\ref{rangesI}). These points in parameter space are 
characterized by small $|A_t|$, which favours small values of 
$B(b\rightarrow s\gamma)$ when $\tan\beta$ is large.}
\label{mike_bsg}
\end{figure}
\begin{figure}
\dofig{6in}{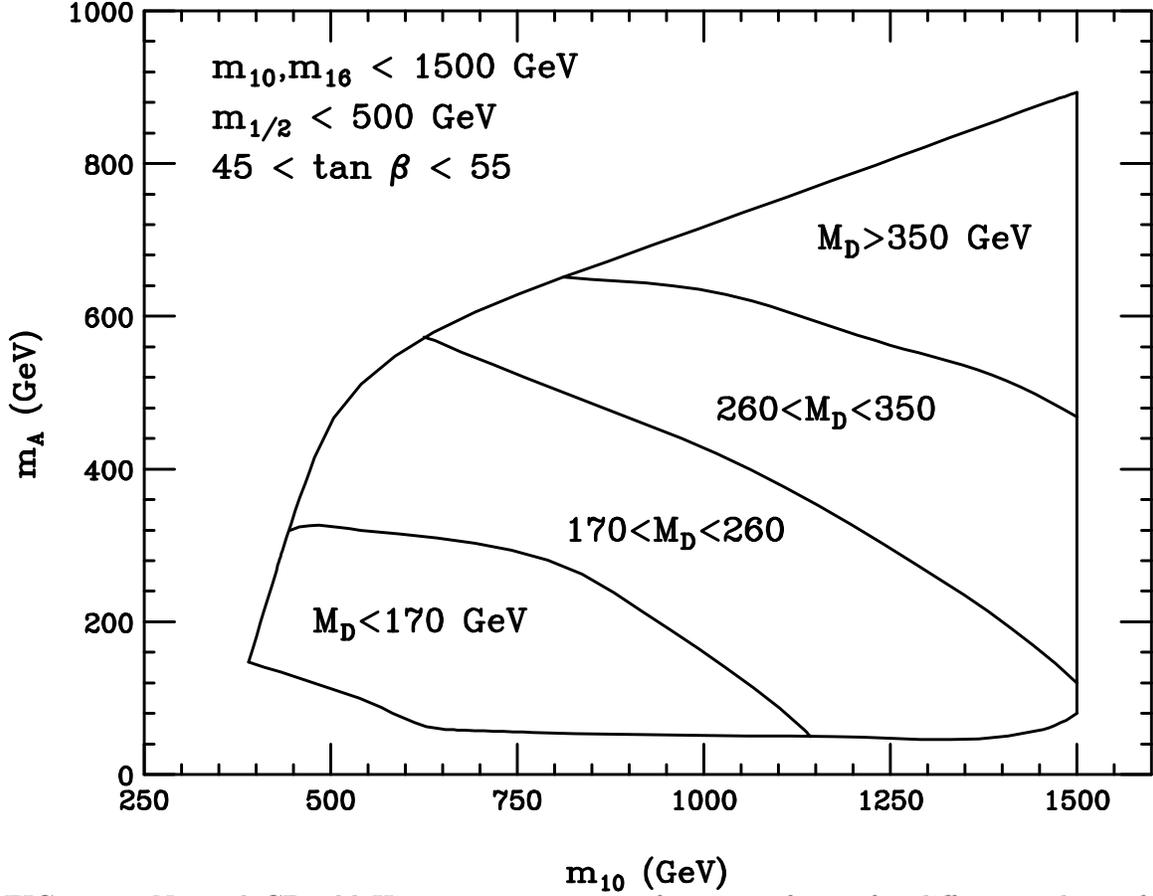}
\caption[]{
Neutral CP-odd Higgs mass $m_A$ as a function of $m_{10}$ for different
values of the D-term parameter $M_D$. The heavy neutral CP-even Higgs $H$
has a mass very close to $m_A$ except when $m_A<100$ GeV.}
\label{mam10_md}
\end{figure}
\begin{figure}
\dofig{7in}{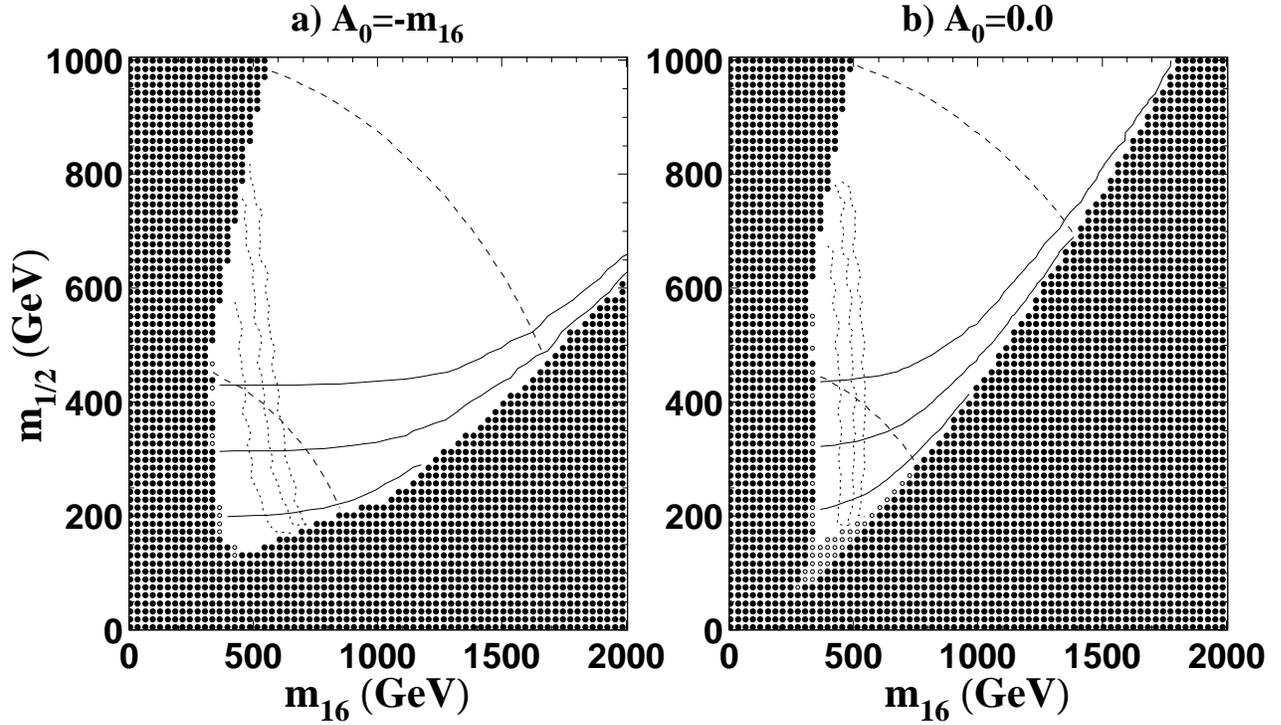}
\caption[]{
The same parameter space planes as in Fig. \ref{pam_yu}, but now showing 
regions of parameter plane where 
$5\sigma$ $b\bar{b}A$ and $b\bar{b}H$ signals
may be seen with the $D0$ detector at Fermilab Tevatron for
(contours from left to right) 
integrated luminosities of 2, 10 and 30 fb$^{-1}$.}
\label{pam_ha}
\end{figure}
\begin{figure}
\dofig{7in}{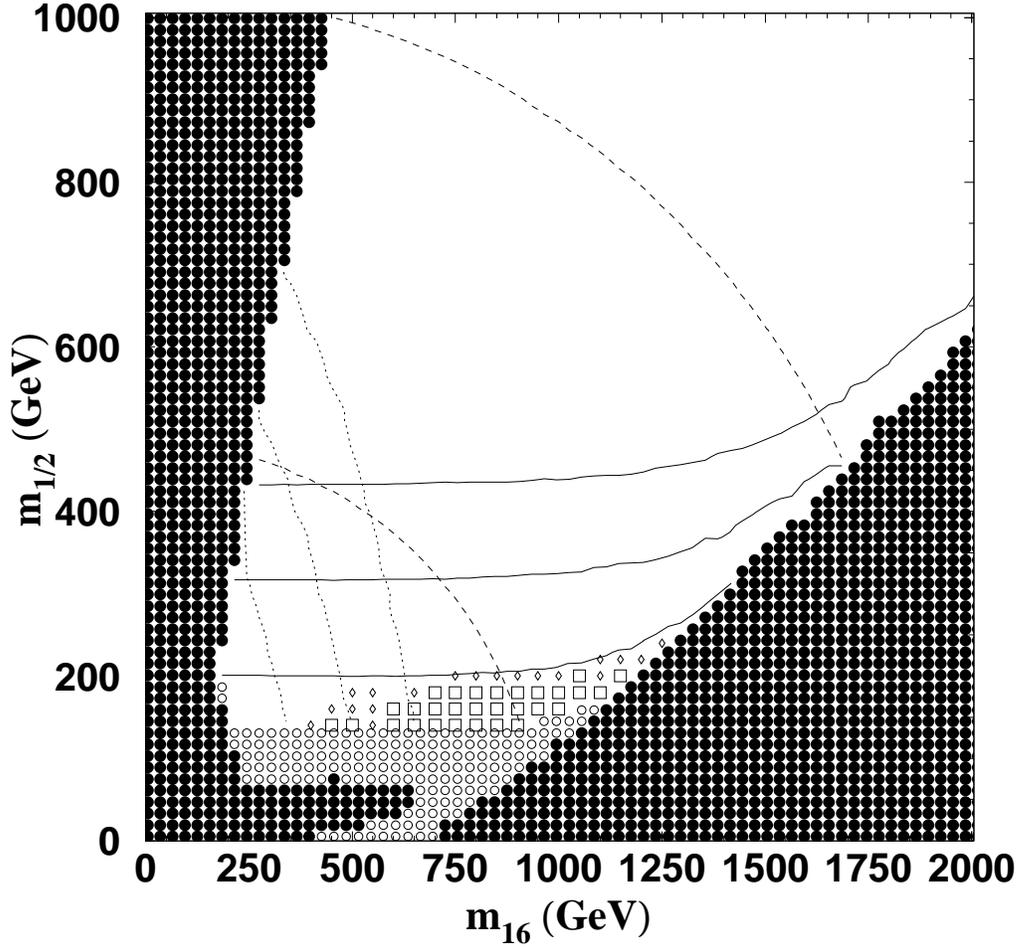}
\caption[]{
Regions of the $m_{16}\ vs.\ m_{1/2}$ parameter plane where trilepton
signals may be seen at the Fermilab Tevatron. Open squares can be
seen at the $5\sigma$ level with 25 fb$^{-1}$ of data, while diamonds can
be seen at the $3\sigma$ level. We take $A_0=0$, $\tan\beta =47$, 
$m_{10}=1.14 m_{16}$, $M_D={1\over 3}m_{16}$ and $\mu <0$.}
\label{pam_3l}
\end{figure}
\begin{figure}
\dofig{6in}{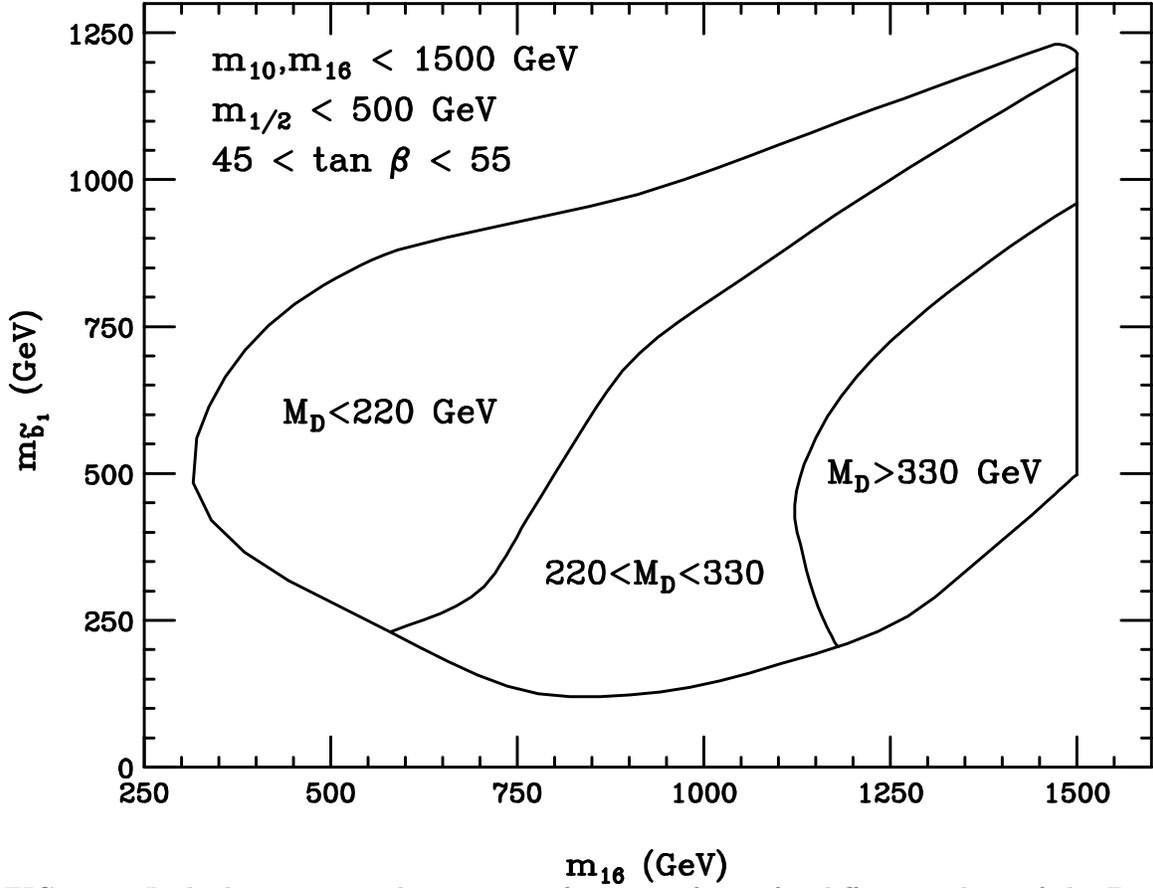}
\caption[]{
Light bottom squark mass as a function of $m_{16}$ for different values of 
the D-term parameter $M_D$. Very light sbottoms can be obtained for 
intermediate values of $M_D$.}
\label{sb1m16_md}
\end{figure}
\begin{figure}
\dofig{7in}{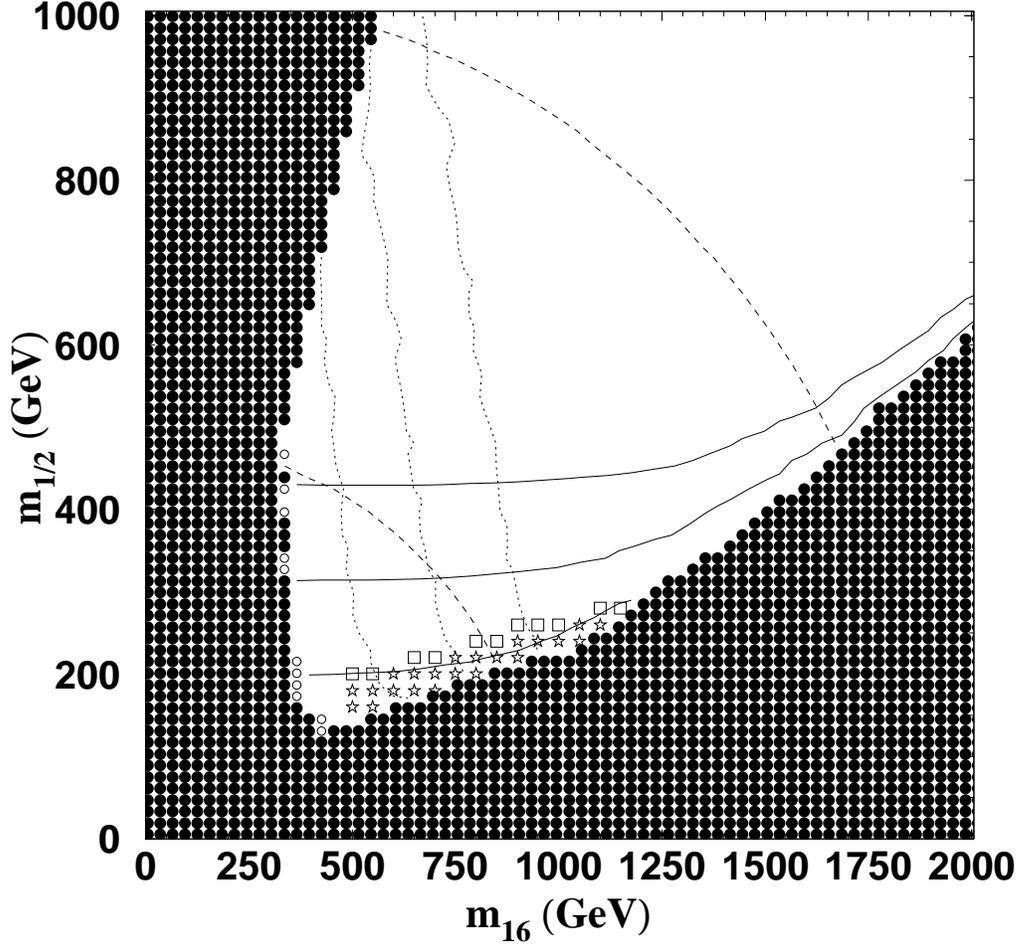}
\caption[]{
Regions of the $m_{16}\ vs.\ m_{1/2}$ parameter plane where 
$\tb_1\bar{\tb}_1\to b\bar{b}+\eslt$ events can be seen above 
background at the Fermilab Tevatron. The stars show points where a 
$5\sigma$ signal can be seen at Run 2 (2 fb$^{-1}$) and the open squares
show points where a $5\sigma$ signal can be seen at Run 3 (25 fb$^{-1}$).
We take $A_0=-m_{16}$, $\tan\beta =50$, 
$m_{10}=1.25 m_{16}$, $M_D={1\over 3}m_{16}$ and $\mu <0$.}
\label{pam_b}
\end{figure}

\end{document}